\documentclass[fleqn,usenatbib]{mnras}
\usepackage{newtxtext,newtxmath}
\usepackage[T1]{fontenc}
\DeclareRobustCommand{\VAN}[3]{#2}
\let\VANthebibliography\thebibliography
\def\thebibliography{\DeclareRobustCommand{\VAN}[3]{##3}\VANthebibliography}

\usepackage{graphicx}	% Including figure files
\usepackage{amsmath}	% Advanced maths commands
\usepackage{xcolor}
\usepackage{tablefootnote}
\usepackage{orcidlink}

\newcommand{\ulas}{ULASJ2315+0143\,}
\title[\ulas]{A Big Red Dot: Scattered light, host galaxy signatures and multi-phase gas flows in a luminous, heavily reddened quasar at cosmic noon} %For now....

\author[M. Stepney et al.]{Matthew Stepney,$^{\orcidlink{0000-0002-7711-0537}\,}$$^{1}$\thanks{E-mail: ms10g17@soton.ac.uk} Manda Banerji,$^{\orcidlink{0000-0002-0639-5141}\,}$$^{1}$ Shenli Tang,$^{\orcidlink{0000-0002-2185-5679}\,}$$^{1}$ Paul C. Hewett,$^{\orcidlink{0000-0002-6528-1937}\,}$$^{2}$ Matthew J. Temple,$^{\orcidlink{0000-0001-8433-550X}\,}$$^{3}$
\newauthor
Clare F. Wethers,$^{4}$ 
Annagrazia Puglisi$^{\orcidlink{0000-0001-9369-1805}\,}$$^{1}$\thanks{Anniversary Fellow.} and 
Stephen J. Molyneux$^{\orcidlink{0000-0003-3596-622X}\,}$$^{1}$
\\
    $^{1}$School of Physics and Astronomy, University of Southampton, Southampton, SO17 1BJ, UK\\
    $^{2}$Institute of Astronomy, University of Cambridge, Madingley Road, Cambridge CB3 0HA, UK\\
    $^{3}$Centre for Extragalactic Astronomy, Department of Physics, Durham University, South Road, Durham DH1 3LE, UK\\
    $^{4}$Department of Space, Earth and Environment, Chalmers University of Technology, Onsala Space Observatory, 439 92 Onsala, Sweden\\  
}

\date{Accepted XXXX. Received XXXX ; in original form XXXX}

\pubyear{2023}

\begin{document}
\label{firstpage}
\pagerange{\pageref{firstpage}--\pageref{lastpage}}
\maketitle

\begin{abstract}

\noindent We present a deep X-Shooter rest-frame UV to optical spectral analysis of the heavily reddened quasar, \ulas at $z=2.566$, known to reside in a major-merger host galaxy. The rest-frame optical is best-fit by a dust-reddened quasar (E(B-V)$^{\rm QSO}= 1.55$) with black-hole mass $\rm log_{10}(H\beta, M\textsc{bh} [M_{\odot}]) = 10.26 \pm 0.05$, bolometric luminosity $\rm L_{Bol}$ = $ \rm 10^{48.16}\, erg\;s^{-1}$ and Eddington-scaled accretion rate log$_{10}$($\rm \lambda_{Edd}) = -0.19$. We find remarkable similarities between \ulas and the high-redshift Little Red Dots (LRDs). The rest-frame UV cannot be explained by a dusty quasar component alone and requires an additional blue component consistent with either a star-forming host galaxy or scattered AGN light. We detect broad high-ionisation emission lines in the rest-UV, supporting the scattered light interpretation for the UV excess. The scattering fraction represents just 0.05\% of the total luminosity of \ulas. Analysis of the mid infra-red SED suggests an absence of hot dust on torus-scales similar to what is observed for LRDs. The obscuring medium is therefore likely on galaxy scales. We detect narrow, blueshifted associated absorption line systems in \ion{C}{iv}, \ion{N}{v}, \ion{Si}{iv} and \ion{Si}{iii}. There is evidence for significant high-velocity (>1000 $\rm km\,s^{-1}$) outflows in both the broad and narrow line regions as traced by \ion{C}{iv} and [\ion{O}{iii}] emission. The kinetic power of the [\ion{O}{iii}] wind is $\dot \epsilon_{k}^{ion} = 10^{44.61} \rm erg\,s^{-1} \sim 0.001\,L_{Bol}$.  \ulas is likely in an important transition phase where star formation, black-hole accretion and multi-phase gas flows are simultaneously occurring. 
\end{abstract}


\section{Introduction}

Key to understanding the evolution of massive galaxies is the study of feedback driven by Active Galactic Nuclei (AGN) powered by the central supermassive black holes in galaxies. Cosmological simulations have demonstrated that feedback processes are pivotal to explaining a multitude of observables, such as the $\rm M - \sigma$ relation \citep{Magorrian_1998,2013ARA&A..51..511K,2021MNRAS.503.1940H}, the shape of the galaxy luminosity and stellar mass function \citep{2001MNRAS.326..255C,Benson_2003,2003ApJ...584..203H,somerville_08,2015MNRAS.446..521S} and the observed morphologies and colours of present-day galaxies \citep{2016MNRAS.463.3948D,morphology_sim}.

It is likely that most massive galaxies have undergone at least one major merger event to account for the mass distribution of massive galaxies in the local Universe \citep{2003AJ....126.1183C,2012A&A...548A...7L}. In such a paradigm of massive galaxy formation, rapid gas inflow to the nucleus fuels a powerful quasi-stellar object (QSO) whilst simultaneously inducing a starburst phase \citep{2005Natur.433..604D,2005ApJ...630..705H,Hopkins_2008}. The starburst galaxies are shrouded in dust, appearing as either ultra-luminous infrared galaxies (ULIRGs) or sub-mm galaxies (SMGs) \citep{Veilleux_2009,2014ApJ...788..125S}. The same gas supply that fuels star formation also fuels accretion onto the central supermassive black hole, which is initially dust obscured \citep{Granato_2004,2005Natur.433..604D,HOP_ELV}. Powerful feedback processes can then couple energy and momentum from the accreting supermassive black hole (SMBH) to the interstellar medium (ISM) quenching star formation during a "blow-out" phase \citep{Fabian:12, Harrison:24} with the presence of dust potentially making radiatively driven outflows during this phase more efficient (e.g. \citealt{2018MNRAS.479.2079C,Ishibashi:17}).  

Red quasars, which show dust extinction affecting the quasar UV continuum have been postulated to correspond to this "blow-out" phase in massive galaxy formation \citep{2008ApJ...674...80U,2012MNRAS.427.2275B,2015MNRAS.447.3368B,2012ApJ...757...51G,2015ApJ...806..218G,2015ApJ...804...27A,2021A&A...649A.102C}. A variety of different methods have been used to select red quasars with varying levels of extinction including optical \citep[e.g.][]{2019MNRAS.488.3109K,2020MNRAS.494.4802F}, near infra-red \citep[e.g.][]{2012ApJ...757...51G,2015MNRAS.447.3368B} and mid infra-red \citep[e.g.][]{Eisenhardt_2012,2015ApJ...804...27A} colour selections. Many red quasars are found to reside in major mergers \citep[e.g.][]{2008ApJ...674...80U,2015ApJ...806..218G,2021MNRAS.503.5583B}. Powerful feedback signatures have also been detected in several of these populations although it is debated whether these feedback processes are more efficient in the obscured quasar phase relative to the unobscured quasar phase. For example, extreme broad line region (BLR) and narrow line region (NLR) outflows have been reported among populations of Extremely Red Quasars (ERQs; \citealt{2017MNRAS.464.3431H,2019MNRAS.488.4126P,2024MNRAS.527..950G}), with their kinetic power exceeding that of blue quasars at comparable luminosities \citep{2016MNRAS.459.3144Z,2019MNRAS.488.4126P}. The correlation between dust extinction and radio emission in the mildly obscured red quasars in SDSS and the even redder quasars in DESI have been interpreted as evidence for low-power radio jets or winds causing shocks in a dusty environment \citep{2023MNRAS.525.5575F}. On the other hand some studies have found very similar ionised gas outflow properties between red and blue quasars when the two samples are matched in luminosity and redshift \citep[e.g.][]{2019MNRAS.487.2594T,Fawcett:22}. Spatially resolved observations of multi-phase outflows in red quasars are also now starting to uncover the complex gas dynamics in these systems (e.g. \citealt{Wylezalek:22,2023ApJ...953...56V}.

Ground-based surveys of the most luminous obscured quasars at $z\sim1-4$ suggest they could dominate the number densities at the highest intrinsic luminosities implying that this population of previously unstudied AGN may account for some of the most luminous and massive accreting black holes at cosmic noon \citep{2015ApJ...804...27A,2015MNRAS.447.3368B}. More recently an abundant population of Little Red Dots (LRDs) observed with \textit{JWST} has emerged, many of which are postulated to be obscured AGN, hundreds of times more common than UV-bright AGN at the faint end of the luminosity function \citep{Onoue_2023,Kocevski_2023,2024arXiv240403576K,2024ApJ...964...39G}. Therefore the obscured phase of supermassive black hole growth also appears to be increasingly important at lower luminosities, and in the high-redshift Universe.

Intriguingly, despite their red colour selection, several of the red quasar populations appear to show excess emission at rest-frame UV wavelengths that is inconsistent with an AGN SED affected by dust extinction \citep[e.g.][]{2015ApJ...804...27A,2018MNRAS.475.3682W,2019ApJ...876..132N,2020ApJ...897..112A,2022ApJ...934..101A,Kocevski_2023,2024ApJ...964...39G,2024ApJ...968...34W,2024ApJ...968....4P,2024ApJ...963..129M,2024arXiv240403576K}. In the population of Hot Dust Obscured Galaxies (HotDOGs) polarisation studies attribute the UV excess to a scattered light component from the heavily cocooned AGN \citep{2020ApJ...897..112A}. In ERQs, the extreme column densities ($\rm N_H \gtrsim 10^{24}\, cm^{-1}$) reported are suggestive of a dense circumnuclear gas cloud suppressing the accretion disk emission \citep{Goulding_2018}. Spectropolarimetry studies of this population also suggest scattering from equatorial dusty disk-winds as the primary cause of their UV emission \citep{2018MNRAS.479.4936A}. Furthermore, $\sim20$ per cent of the broad-line AGN recently observed with \textit{JWST} are both obscured and feature relatively blue colours in the rest-frame UV, with many more unconfirmed AGN candidates demonstrating similar photometry \citep{Kocevski_2023,2024ApJ...963..129M,2024ApJ...964...39G,2024arXiv240403576K}. The broadband UV excess can be explained by a scattered light component or emission from unobscured star-forming regions in the AGN host galaxy \citep{2024ApJ...964...39G} and spectroscopic observations would help to discriminate between these scenarios. 

In this paper we conduct a detailed study of a heavily reddened quasar, \ulas at $z\sim2.56$. Heavily reddened quasars (HRQs) are one class of red quasars selected based on their red near infra-red colours. Over 60 HRQs have been selected using near and mid-infrared colour selections using data from the UKIDSS, VISTA and \textit{WISE} surveys \citep{2012MNRAS.427.2275B,2013MNRAS.429L..55B,2015MNRAS.447.3368B,2019MNRAS.487.2594T}. They cover a redshift range $0.7 \lesssim z \lesssim 2.7$ and have measured dust extinctions of $0.5 \lesssim$ E(B-V) $\lesssim 3.0$. In the X-ray, HRQs are amongst the most powerful quasars known, with a median $\rm \langle L_{2-10kev}\rangle=10^{45.1} \,erg\,s^{-1}$ \citep{2020MNRAS.495.2652L}. As expected from the heavy attenuation observed in the rest-frame optical, HRQs have large gas column densities $\rm N_H \sim 10^{22-23}\,cm^{-2}$, suggestive of a blow-out phase given their high Eddington-scaled accretion rates \citep{2020MNRAS.495.2652L}. Atacama Large Millimeter/submillimeter Array (ALMA) observations have revealed diverse gas fractions, gas morphologies and ISM properties in HRQs \citep{Banerji:17,2018MNRAS.479.1154B,2021MNRAS.503.5583B}. 
Finally, deep broad-band optical photometry of a sub-sample of 17 HRQs reveals that at least 10 feature blue colours in the rest-frame UV inconsistent with the dust-reddened quasar SED in the rest-frame optical. The UV-excess potentially originates from relatively unobscured star-forming host galaxies with average $\rm SFR_{UV} = 130\pm95\, \emph{M}_{\odot}\, yr^{-1}$ \citep{2018MNRAS.475.3682W}. However, spectroscopic data in the rest-frame UV is required to confirm this hypothesis. 

We selected \ulas for follow-up with X-Shooter due to its detection at rest-frame UV wavelengths across multiple bands in the HyperSuprimeCam (HSC) survey as well as the rich array of multi-wavelength observations already assembled for this source. X-ray data suggests that the inner regions of \ulas are dominated by a powerful accreting SMBH ($L_X = 3 \times 10^{45}\,\rm erg\,s^{-1}$) with moderately high gas column densities ($N_H \sim 10^{22}\,\rm cm^{-2}$), making it the ideal candidate for studying radiation-driven winds \citep{2020MNRAS.495.2652L}. At longer wavelengths, \ulas has been confirmed to be undergoing a gas-rich major merger via ALMA observations of the warm molecular gas as traced by CO(3-2) \citep{2021MNRAS.503.5583B}. Further CO(1-0) observations from the Very Large Array (VLA) confirm that the merging galaxies reside within an extended reservoir of cold molecular gas, spatially offset from the merging pair and redshifted with respect to the CO(3-2) observations \citep{2018MNRAS.479.1154B}. In addition, ALMA observations of the dust continuum suggest a total star formation rate of $\sim 680\, M_\odot\,\rm yr^{-1}$, consistent with a merger-induced starburst. In the rest-frame optical, \ulas is amongst the reddest of the HRQ sample and hence exhibits the most unambiguous evidence of excess rest-frame UV emission \citep{2018MNRAS.475.3682W}. Owing to its broad wavelength coverage, the analysis of the X-Shooter spectrum will enable the detailed study of \ulas across the entire rest-frame UV to optical wavelength region, leading to a better understanding of the UV excess emission and probing key diagnostic emission and absorption features that enable the study of multi-phase and multi-scale winds. 

The paper is organised as follows: Section \ref{SEC:Reduction} details the observations and reduction of the X-Shooter spectrum. In Section \ref{SEC:Results} we model the spectral energy distribution and analyse the emission and absorption line properties. We discuss our results in Section \ref{SEC:DISC} in the context of other red AGN populations. Our conclusions are presented in Section \ref{SEC:Conc}. Vacuum wavelengths and AB magnitudes are employed throughout the paper and we adopt a $\Lambda \textnormal{CDM}$ cosmology with $h_0$ = 0.71, $\Omega_\textnormal{M}$ = 0.27 and $\Omega_\Lambda$ = 0.73.

\section{Spectral Observations \& Data Reduction} \label{SEC:Reduction}

We observed \ulas (RA=348.9842, Dec=1.7307) with the X-Shooter spectrograph on the Very Large Telescope (VLT) (see Data Availability for access to the raw data). X-Shooter is an Echelle spectrograph installed on the UT2 telescope at the VLT \citep{2011A&A...536A.105V}. It is comprised of three arms; the Ultra-Violet/Blue (UVB), the Visible (VIS) and the near-infrared (NIR). The combined wavelength coverage of X-Shooter ranges from $\sim$3100-24000\AA,\,enabling the study of both the rest-frame UV and rest-frame optical emission of \ulas at $z\sim2.5$. The spectra were obtained using a 2x2 binning to minimise read-out noise. Slit widths were set to 1.3 arc-seconds for the UVB and 1.2 arc-seconds for both the VIS and NIR arms, yielding a spectral resolution $\rm R=$ 4100, 6500, 4300, for the UVB, VIS and NIR arms, respectively. The target was observed for $\sim$9 hours based on its broadband SED \citep{2018MNRAS.475.3682W} and to ensure sufficient signal-to-noise ratio (S/N) across the entire wavelength range covered by X-Shooter. To aid scheduling, these observations were split into roughly one hour long observing blocks and a standard ABBA nodding pattern was adopted for the purposes of sky subtraction in the NIR arm. 

We employ PypeIt v1.14.0\footnote{\url{https://github.com/pypeit/PypeIt/tree/1.14.0}},
a Python package for semi-automated reduction of astronomical slit-based spectroscopy \citep{pypeit:zenodo,pypeit:joss_pub}, for the spectral reduction. Some user-level parameters had to be adjusted from their default values and these are detailed in Appendix \ref{APP:PypeIt_User_Params}. Wavelength calibrations in the X-Shooter UVB/VIS arms were conducted by matching arc lines from Thorium-Argon (ThAr) lamps to the inbuilt ThAr arc line catalogue stored within PypeIt. For the NIR arm, wavelength calibrations were conducted using the OH sky line transitions from the spectrum itself. Pypeit's extraction algorithm is based on \citet{1986PASP...98..609H}. A third order polynomial fit is attempted across all orders in a given exposure. Echelle objects are retained if the S/N of the trace exceeds a certain threshold value in all orders, or if the S/N of the trace exceeds the minimum S/N threshold over a user-defined number of orders (See Appendix \ref{APP:PypeIt_User_Params} for details). The standard star trace is used as a crutch for the polynomial fit for those orders that do not exceed the minimum S/N requirement. The relative flux calibration was conducted within PypeIt, and the spectrum was then normalised to the cModel magnitude of the quasar from Hyper Suprime-Cam (HSC) in the $g$, $i$ and $z$ bands\footnote{Calibration to the Dark Energy Camera photometry for this source yields $\Delta g = 0.06 \rm \,mag$ \citep{2018MNRAS.475.3682W}, hence we estimate an $\sim6\%$ uncertainty in the flux calibration}, and the United Kingdom Infrared Deep Sky Survey (UKIDSS) magnitudes in the $H$ and $K$-bands \citep{2015MNRAS.447.3368B}.

\begin{figure*}
 \includegraphics[scale=0.8]{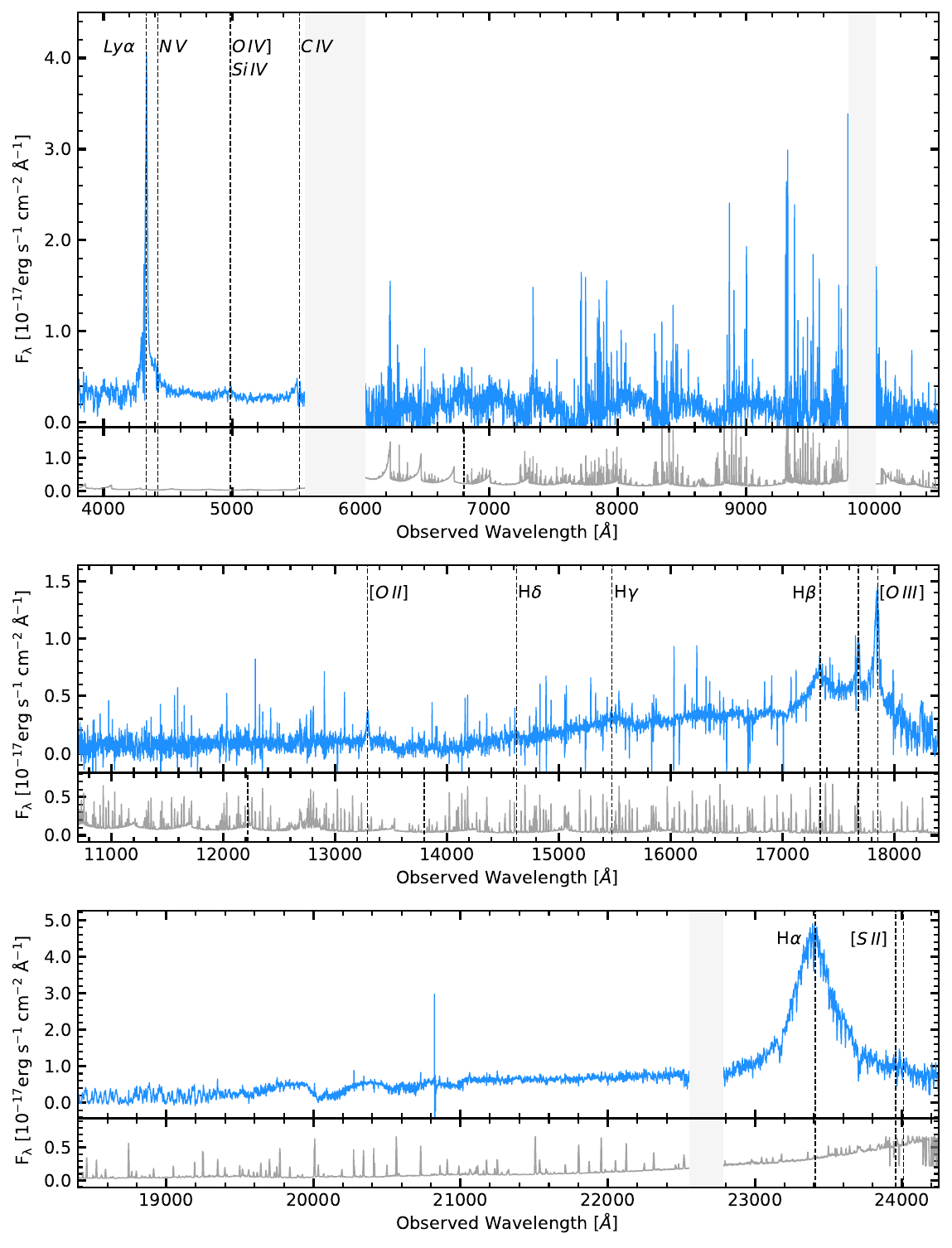} 
  \caption{The full flux calibrated 1D spectrum of \ulas is presented in the top panels of each exert (blue). In addition, the corresponding noise spectrum is presented in the lower panels (grey). Noise spikes marking the transition between arms at $\sim$ 5800\AA\, and $\sim$ 10,000\AA\, have been masked out, as too has a region of high noise redward of the H${\alpha}$ emission. Prominent emission lines are marked by dashed vertical black lines and labelled accordingly.}
 \label{fig:Reduced_Spec_Complete}
\end{figure*}

\autoref{fig:Reduced_Spec_Complete} depicts the full flux calibrated 1D X-Shooter spectrum of \ulas, as well as the corresponding noise spectrum. As expected from the red NIR colour of \ulas ($(H - K) > 1.4$; \citealt{2015MNRAS.447.3368B}), the NIR spectral shape (middle/bottom; Fig.\ref{fig:Reduced_Spec_Complete}) is consistent with significant dust attenuation toward the quasar continuum. The median continuum S/N in the intervals [16000,17000]\AA\, and [21000,22000]\AA\, are $\sim 13$\AA$^{-1}$ and $\sim 12$\AA$^{-1}$, respectively. We observe narrow emission lines, including [\ion{O}{ii}]\,$\lambda\lambda3726,3729$ and [\ion{O}{iii}]\,$\lambda\lambda4960,5008$ in the NIR spectrum. However, there is no discernible signal from either [\ion{Ne}{iii}]\,$\lambda3869$ or [\ion{Ne}{v}]\,$\lambda3426$. We see several broad Balmer lines in the NIR, including the H$\gamma\,\lambda4342$, H$\beta\,\lambda4863$ and H$\alpha\,\lambda6565$ emission. The median continuum S/N in the interval [6000,10000]\AA\, is $\sim 1$\AA$^{-1}$ (top/right; Fig.\ref{fig:Reduced_Spec_Complete}). When rebinned by a factor of ten, the median continuum S/N $\sim 0.2$\AA$^{-1}$, hence the flux in this region is dominated by noise. The median continuum S/N in the interval [4600,4900]\AA\, (top/left; Fig.\ref{fig:Reduced_Spec_Complete}) is $\sim 15$\AA$^{-1}$. The rest-frame UV emission is inconsistent with the attenuation observed at redder wavelengths, which would predict a much lower UV flux. We see broad Ly$\alpha$, \ion{N}{v} and \ion{C}{iv} emission features on a blue UV continuum. We also observe a strong narrow component to the Ly$\alpha$ emission as well as absorption around the Ly$\alpha$ and \ion{C}{iv} emission line profiles. These spectral features are fully explored in Section \ref{SEC:Line_properties}.

\section{Results} \label{SEC:Results}
\subsection{Rest-frame UV Image Decomposition}

\begin{figure*}
 \includegraphics[scale=0.5]{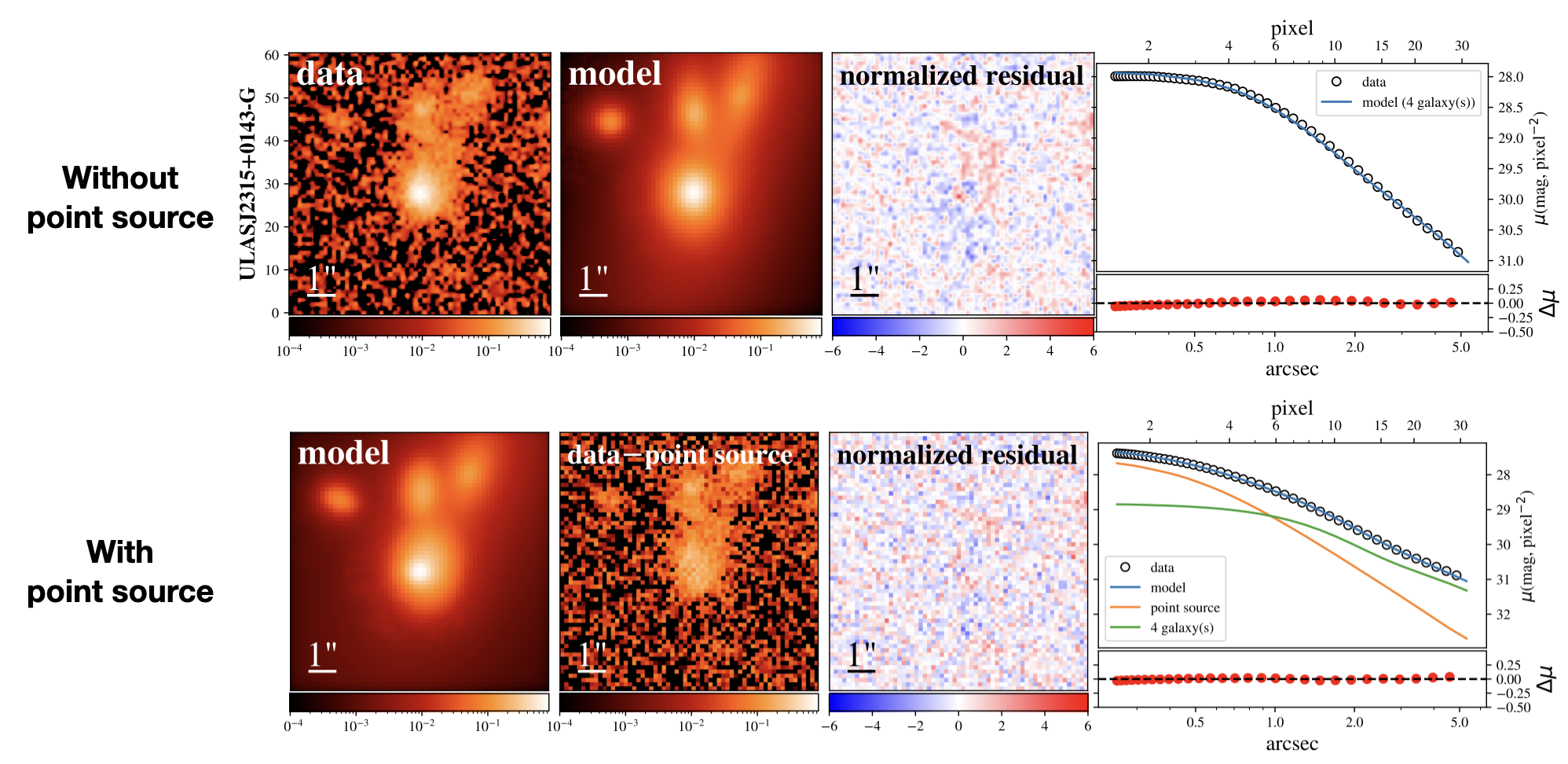} 
  \caption{\textsc{GaLight} fitting results on the HSC $g$ band image of \ulas. The first row shows the case when use a model incorporating only the S\'ersic profiles. The second row shows the case when we include one point source in the model. The reconstructed images are shown by the second and first panel of the rows, respectively. In the second panel of second row, we also show the image after subtracting the point source model from the data. The third column shows the 2D normalised residual map of the fitting. The final column shows the 1D annulus light profiles for the data and the reconstructed model, with the fractional residual shown in the bottom sub-panel.}
 \label{fig:galight}
\end{figure*}

We begin by conducting image decomposition analysis to better understand the source of the rest-frame UV emission in \ulas. We consider the morphology of this emission as traced by the imaging data from HSC in the $g$-band. We follow the work of \citet{silverman2020dual} and \citet{tang2021optical} to perform a 2D image decomposition analysis on the HSC $g$-band image of \ulas, which covers 4000-5500\AA\, in the observed frame \citep{kawanomoto2018hyper}. We apply the tool \textsc{GaLight} \citep[Galaxy shapes of Light][]{ding2022galight}, which was developed based on the image modelling capabilities of \textsc{lenstronomy} \citep{2021JOSS....6.3283B}. The tool accepts a preset of number of point sources to be used in fitting. We tested three cases: (i) no point source (i.e. no AGN contribution to the rest-frame UV), (ii) one point source (i.e. a single quasar contribution to the rest-frame UV), and (iii) two point sources (i.e. a dual quasar contribution to the rest-frame UV). As a result, the algorithm failed to find a secondary point source in the third case, and favoured at most one point source in the region. The algorithm then finds as many extended sources as necessary in the field, and fits them with S\'ersic profiles simultaneously with the point sources. We show the results in Fig. \ref{fig:galight}, with the first row showing the case of no point source and the second row showing the model for one point source. For the model with one point source, we also show the image after subtracting the point source model from the data (second row, second column). The third column shows the normalised residual of the fitting. The final column shows the 1D annulus light profiles from the centre of the source to the outer regions. The reconstructed models are shown by the blue curves, in comparison with the data, shown by the open circles. The bottom sub-panels show the fractional residuals in 1D.

We find that the light profile of this system can either be explained by one point-spread function (PSF) and four S\'ersic profiles or just four S\'ersic profiles. We also compared the reduced $\chi^2$ of the two fits, which are 1.09 for the model without a point source, and 1.04 for the model with a point source. We therefore conclude that the rest-frame UV morphology of this source can be equally well reconstructed with and without a PSF contribution to the $g$-band. We are therefore motivated to explore spectral energy distribution (SED) models that exploit both star-forming host galaxy emission or blue quasar emission.

\subsection{Spectral energy distribution} \label{SEC:SED_Fit}

In this section we explore the rest-frame UV to optical SED of \ulas. As noted from Fig. \ref{fig:Reduced_Spec_Complete}, the spectral shape is inconsistent with a single reddened quasar SED with excess blue flux detected in the UVB arm relative to this SED. Motivated by Fig.\ref{fig:galight}, we explore two distinct possibilities to explain the excess UVB flux - (i) a blue quasar component either originating from leaked/scattered light or a secondary AGN and (ii) a star-forming host galaxy. We model the quasar components using the \textsc{qsogen}\footnote{\url{https://github.com/MJTemple/qsogen}} tool, a Python package that implements an empirically-motivated parametric model to simulate quasar colours, magnitudes and SEDs \citep{2021MNRAS.508..737T}. The host galaxy component is modelled using the \textsc{bagpipes}\footnote{\url{https://github.com/ACCarnall/bagpipes}} package. Model fits were conducted using a Bayesian Markov-Chain Monte-Carlo (MCMC) method. We utilise \textsc{emcee}, a Python package\footnote{\url{https://github.com/dfm/emcee}} that explores the likelihood space using the affine-invariant ensemble sampler proposed by \citet{2010CAMCS...5...65G}.  

\subsubsection{Reddened quasar + blue quasar light}

\begin{figure*}
 \includegraphics[scale=0.75]{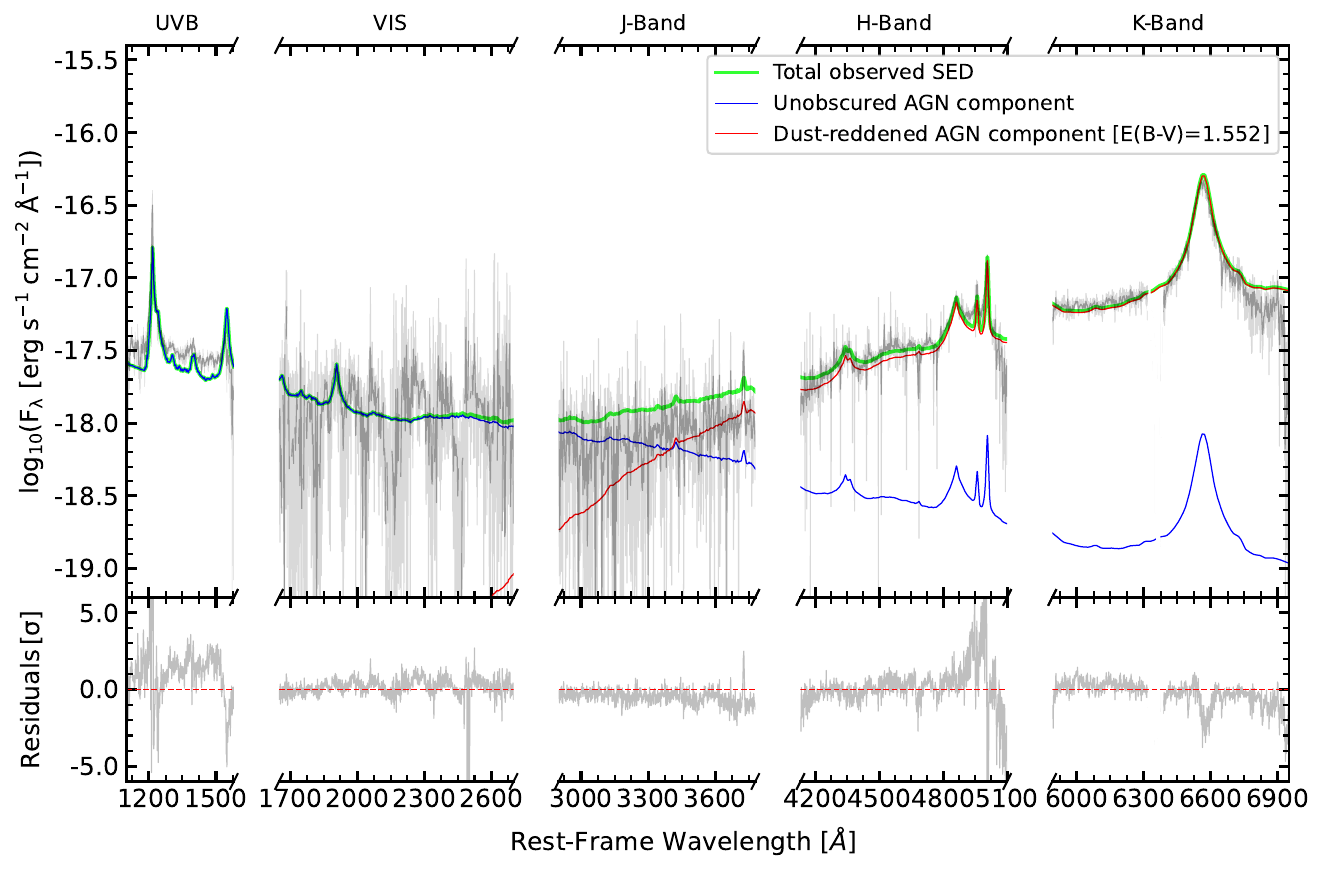} 
  \caption{We present the spectrum of \ulas in light grey with the 20-pixel rebinned spectrum overlaid in dark grey. Our best-fit SED model is plotted in green. The unobscured AGN component is presented in blue, while the dust-reddened AGN component is presented in red. Regions of high telluric absorption have been masked to aid readership. The bottom panel illustrates the model residuals.}
 \label{fig:Temple_SED_Fit}
\end{figure*}

The first model explored is the combination of a reddened quasar SED with some blue AGN component representing a fraction of the total quasar luminosity leaked/scattered along our line of sight or a less luminous secondary AGN. The free parameters are the luminosity of the primary quasar, log$_{10}\{\lambda$L$_{\lambda} (3000\mathring{A}) [\rm erg\;s^{-1}]\}$, the dust reddening of the primary quasar, E(B-V)$^{\rm QSO}$ and the fraction of the total intrinsic SED of the primary quasar emitted in the rest-frame UV, $f_{UV}$. \textsc{qsogen} also allows flexibility in the emission line contributions to the quasar SED via the \emph{$emline\_type$} parameter, which essentially controls the equivalent widths of the key quasar emission lines \citep{2021MNRAS.508..737T}. Hence, we also set the \emph{$emline\_type$} as a free parameter in the fit. The model includes no contribution from the quasar host galaxy to the SED. The quasar extinction law assumed by \textsc{qsogen} is discussed in Section 2.6 of \citet{2021MNRAS.508..737T} and is similar to those derived by \citet{2004MNRAS.348L..54C} and \citet{2010A&A...523A..85G}. We permit the \textsc{emcee} package to explore an N-dimensional Gaussian likelihood function, where N represents the number of free parameters in the fit (N=4), and apply uniform priors. Regions significantly affected by telluric absorption are excluded from the SED fit. The best-fit SED is presented in Fig. \ref{fig:Temple_SED_Fit} overlaid on the X-Shooter spectrum.

The model is able to successfully reproduce the spectrum redward of $\sim$4200\AA\, as well as the the broad shape of the SED blueward of 1600\AA.\, However in the interval [2900,3800]\AA\, the continuum is overestimated. The best-fit parameters of this model are presented in Table \ref{tab:SED_Best_fit} (See Appendix \ref{APP:SED_Corner} for a discussion on model parameter degeneracies), the quoted uncertainties represent the Monte-Carlo uncertainties on the free parameters and hence do not account for the uncertainties associated with the flux calibration. The reddening of the quasar is estimated to be E(B-V)$^{\rm QSO}$$=1.552\pm0.002$, which is consistent with the results from SED-fitting to the broadband photometry conducted by \citet{2018MNRAS.475.3682W}. The dust-corrected optical luminosity is estimated to be log$_{10}\{\lambda$L$_{\lambda} (3000\mathring{A}) [\rm erg\;s^{-1}]\}$$\,=47.907\pm0.003$. Only $0.0450\pm0.0003$ per cent of this intrinsic quasar luminosity is required to be scattered or leaked into our line-of sight to reproduce the UVB flux. This is a much smaller fraction than is reported in other obscured AGN populations at similar redshifts e.g. \citep[$\sim$1-3 per cent in HotDOGs;][]{2020ApJ...897..112A}. We will discuss these results further in Section \ref{SEC:DISC_UV_CONT}.

\begin{table}
    \centering
    \caption{Best-fit \textsc{qsogen} parameters for the SED fit shown in Fig. \ref{fig:Temple_SED_Fit}. Uncertainties represent the MCMC uncertainties.}
    
    \begin{tabular}{|l|l|}
         \hline
         \hline
         Parameter & Best-Fitting Value  \\
         \hline
          emline$\_$type & $1.71\pm0.01$  \\ 
          E(B-V)$^{\rm QSO}$ [mag] & $1.552\pm0.002$  \\ 
          log$_{10}\{\lambda$L$_{\lambda} (3000\mathring{A}) [\rm erg\;s^{-1}]\}$ & $47.907\pm0.003$  \\
          f$_{\rm{UV}}$ & $0.0450\pm0.0003\%$  \\
         \hline
         \hline
    \end{tabular}
    \label{tab:SED_Best_fit}
\end{table}

The primary weakness of this SED model is that it is unable to reproduce the narrow [\ion{O}{ii}] emission as well as over-predicting the continuum flux in the interval [2900,3800]\AA.\, Strong [\ion{O}{ii}] emission in the absence of [\ion{Ne}{iii}] and [\ion{Ne}{v}] could be attributed to ongoing star formation \citep{2018MNRAS.480.5203M}. Despite the need for an AGN component to explain the broad UV emission lines, it is unclear to what extent the AGN is contributing to the continuum. 

\subsubsection{Reddened quasar + star-forming host galaxy} \label{SEC:host+QSO}

The second model explored is then the combination of a star-forming host galaxy and dust reddened quasar SED. We model the host galaxy spectrum using \textsc{bagpipes}. We assume a burst-like star formation history (SFH), consistent with the observation of a gas-rich major merger in the \ulas system \citep{2021MNRAS.503.5583B}. Our goal is to find whether a reasonable star-forming galaxy SED model can reproduce the observed UV-continuum shape. Given the dominance of the quasar emission in the rest-frame optical, we do not necessarily expect to obtain robust constraints on the host galaxy properties. 

We set the duration of the burst, the stellar mass of the galaxy, log(M$_\ast$/M$_\odot$), the galaxy dust extinction A$_{\rm{V}}^{\rm{Gal}}$, the metallicity, log$_{10}$(Z/Z$_\odot$) and ionisation parameter, log$_{10}$(U) as free parameters in the MCMC simulation.  We set flat priors on the metallicity and ionisation parameter in the range $\rm -2<log_{10}(Z/Z_\odot)<0.3$ and $\rm -3.5<log_{10}(U)<-1$. We assume the \citet{2000ApJ...533..682C} extinction law. Finally, we set Gaussian priors on the optical luminosity of the quasar, quasar emission line type and quasar dust extinction, defining the position and width of the priors by the best-fit values and uncertainties of these parameters from Table \ref{tab:SED_Best_fit}. 

\begin{figure*}
 \includegraphics[scale=0.75]{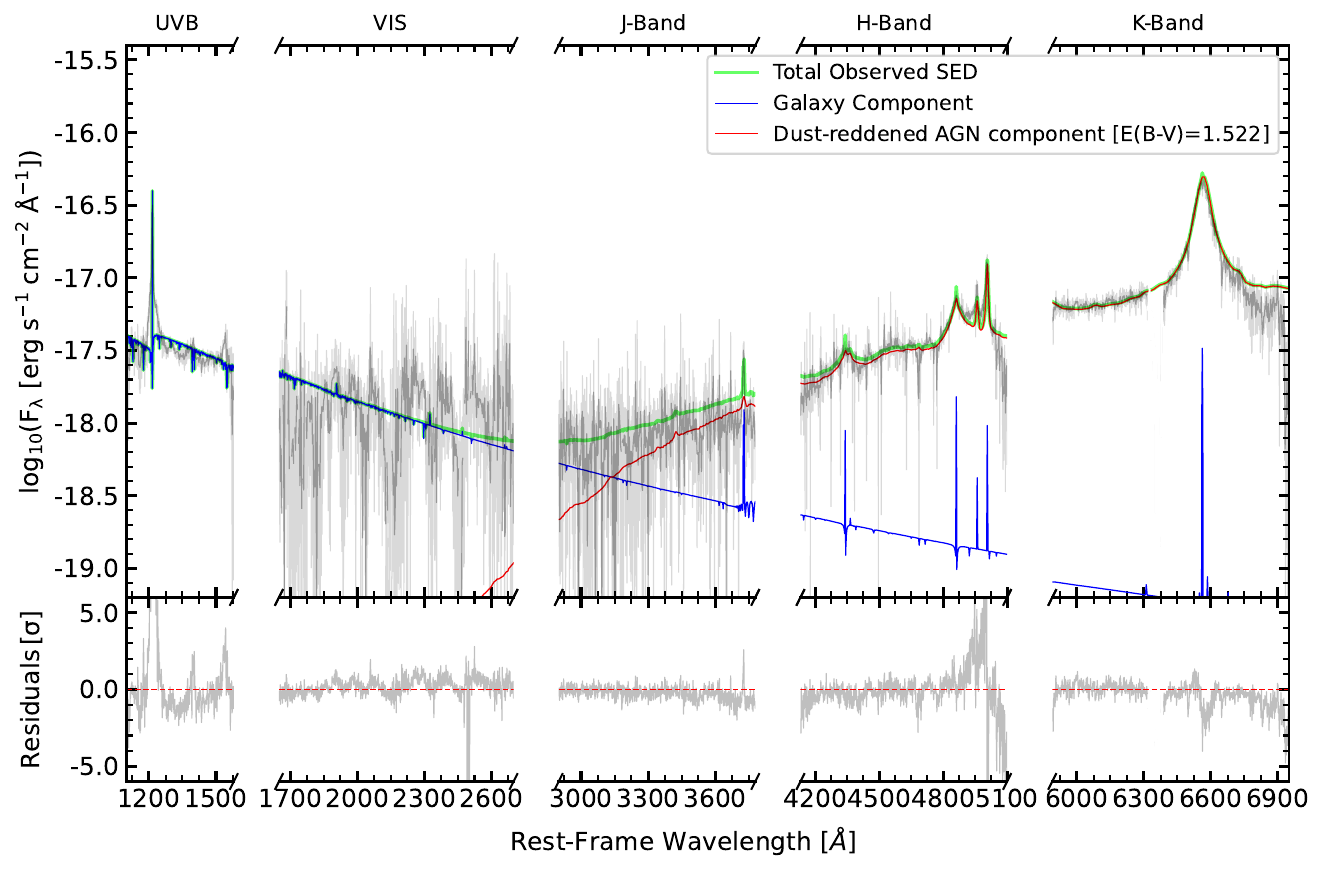} 
  \caption{Same as Fig.\ref{fig:Temple_SED_Fit}, but instead the rest-frame UV is modelled by a star-forming galaxy, generated using \textsc{bagpipes}.}
 \label{fig:Temple_SED_Fit_GAL}
\end{figure*}

The results are presented in Fig.\ref{fig:Temple_SED_Fit_GAL}. We are able to better reproduce the narrow [\ion{O}{ii}] emission and broad continuum shape in the interval [2900,3800]\AA\, with the inclusion of a star-forming galaxy to the SED model. While the star-forming galaxy component is unable to reproduce the broad emission lines observed blueward of 1600\AA,\, the broad continuum shape at these bluer wavelengths is also well fit by the star-forming galaxy. The best-fit parameters for this SED model are presented in Table \ref{tab:SED_Best_fit_GAL_ONLY}, again the uncertainties presented represent the Monte-Carlo errors only, and do not account for uncertainties in the flux calibration. 

\begin{table}
    \centering
    \caption{\textsc{qsogen} and \textsc{bagpipes} user-level parameters for the SED fit shown in Fig. \ref{fig:Temple_SED_Fit_GAL}. Uncertainties represent the MCMC uncertainties.}

    \begin{tabular}{|l|l|}
         \hline
         \hline     
         Parameter & Best-Fitting Value  \\
         \hline        
          emline$\_$type & $1.24\pm0.02$  \\
          E(B-V)$^{\rm QSO}$ [mag] & $1.522\pm0.003$  \\
          log$_{10}\{\lambda$L$_{\lambda} (3000\mathring{A}) [\rm erg\;s^{-1}]\}$ & $47.901\pm0.004$  \\ 
          \hline
          A$_{\rm V}^{\rm Gal}$ [mag] & $0.41\pm0.03$  \\
          log$_{10}$ (M$_\ast$ / M$_\odot$) & $9.47\pm0.02$  \\ 
          t$\rm _{Burst}$ [Myrs] & $10.4\pm0.7$  \\
          $\rm log_{10}(U)$ & $>-1.001\pm0.007$  \\   
          $\rm log_{10}(Z/Z_\odot)$ & $1.70\pm0.04$  \\          
         \hline
         \hline    
    \end{tabular}
    \label{tab:SED_Best_fit_GAL_ONLY}
\end{table}

Firstly, we note that the quasar dust attenuation and continuum luminosity for this model are consistent with the values predicted by the previous SED model, confirming that the continuum at redder wavelengths is indeed quasar dominated. We do not place particular emphasis on the physical values of the galaxy parameters in the fit given the degeneracies between these parameters. The lack of data straddling the 4000\AA\, break, which falls within the gap between the $J$ and $H$-bands, prevents us from placing strong constraints on the stellar mass of the host galaxy and the value of the stellar mass from the SED fit should therefore not be interpreted as representing the total stellar mass of the quasar host. The best-fit ionisation parameter of $\rm -log_{10}(U) = 1.001\pm0.007$ is high for a star-forming galaxy \citep{2018MNRAS.479.2079C}, but can almost certainly be explained by some AGN contamination to the emission lines, many of which have broad components. Nevertheless, we can conclude that a star-forming galaxy SED can reproduce the UV continuum shape and narrow emission line features seen in the X-Shooter spectrum. 

If we assume that the entire rest-frame UV continuum originates from young stellar populations, we can calculate a maximum SFR for \ulas using Eqn. \ref{eq:SFR_L1550};

\begin{equation}
    \begin{aligned}
     \rm SFR_{FUV}(\emph{M}_{\odot}\, yr^{-1}) = log_{10}(\lambda L_\lambda(1550\,\mathring{A})) + log_{10}(C_{FUV}) \\
    \end{aligned}  
    \label{eq:SFR_L1550}    
\end{equation}

\noindent where the continuum flux at 1550\AA,\, $\lambda L_\lambda(1550\,\mathring{A}) = 10^{44.8} \rm erg\,s^{-1}$ and $\rm log_{10}(C_{FUV})$ = 43.35 \citep{2011ApJ...737...67M,2011ApJ...741..124H,2012ARA&A..50..531K}. Eqn. \ref{eq:SFR_L1550} yields $\rm SFR_{FUV} = 88\, \emph{M}_{\odot}\, yr^{-1}$, consistent with the SFR estimated for this object using broad-band photometry \citep[$130\pm95\,\emph{M}_{\odot}\, yr^{-1}$;][]{2018MNRAS.475.3682W}.

\subsubsection{Reddened quasar + blue quasar light + star-forming host galaxy}

Figures \ref{fig:Temple_SED_Fit} $\&$ \ref{fig:Temple_SED_Fit_GAL} show that both an AGN SED and a star-forming galaxy SED can reproduce the broad continuum shape of the UVB spectrum. With both strong narrow [\ion{O}{ii}] in the $J$-band and broad \ion{C}{iv} and \ion{N}{V} in the UVB arm a model that combines a star-forming host galaxy with a blue quasar component to reproduce the flux seems well motivated. This is also justified by the image decomposition in Fig. \ref{fig:galight}, which suggests that both point-like and extended emission contribute to the rest-frame UV flux of this source. 

However, the parameters dictating the nebular emission line strength, such as metallicity, ionisation parameter and age, are degenerate with the stellar mass. A less massive star-forming galaxy with strong nebular emission makes a similar contribution to the total SED as a more massive galaxy with weaker nebular emission. Furthermore, the stellar mass of the galaxy is degenerate with the fraction of the primary quasar contribution to the UV flux, $f_{UV}$. Consequently, with all three components of the SED, the fit failed to converge tending to the limits set by the priors on the stellar mass, metallicity and ionisation parameter. We therefore did not explore a three component SED model further but conclude that the observed spectrum and UV image analysis suggests contributions from both a star-forming host galaxy and a blue AGN component in the UV. 

\subsection{Emission and absorption line properties}\label{SEC:Line_properties}

We now turn our attention to analysing the spectral line properties. We utilise the PyQSOFit package\footnote{\url{https://github.com/legolason/PyQSOFit}}, a python code developed by \citet{2018ascl.soft09008G} for all line fits. In all cases uncertainties on line properties are derived by sampling 500 spectra perturbed by Gaussian noise consistent with the noise array and looking at the mean properties and scatter on these properties from the simulated spectra. 

\subsubsection{Systemic redshift and star formation rate from [\ion{O}{ii}]}

We begin with an analysis of the NIR arm, which traces the rest-frame optical emission in this object. We fit the $J$, $H$ and $K$ band spectra separately so that the continuum fitting is not affected by telluric absorption. 

\begin{figure}
 \includegraphics[width=\linewidth]{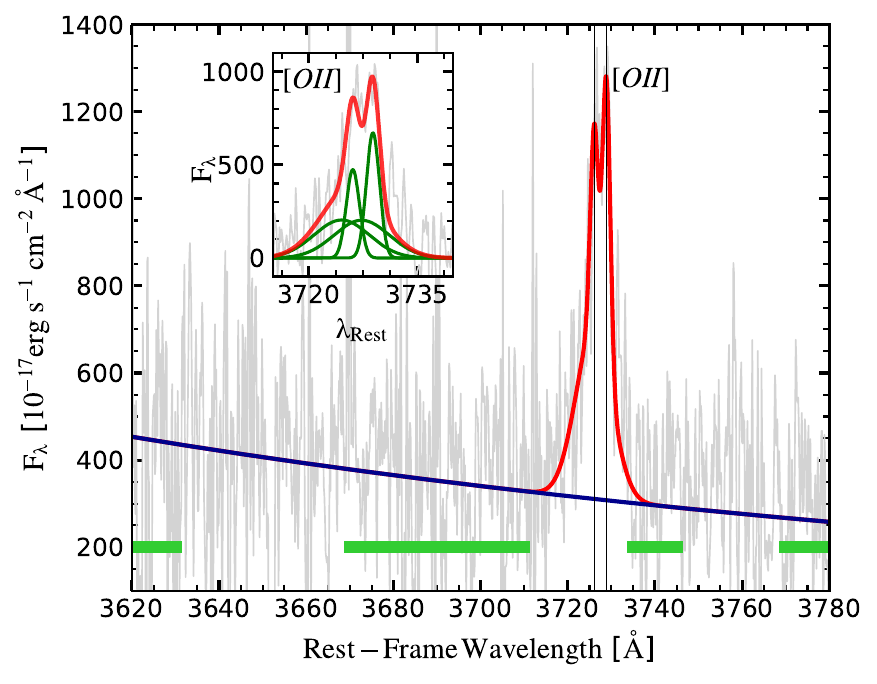} 
  \caption{$J$-Band spectrum for \ulas with a dust correction, E(B-V)$^{\rm QSO}$ = 1.55, applied in grey and showing the [\ion{O}{ii}] doublet. The continuum is presented in navy blue, and the windows over which the continuum was fitted are denoted by the thick, lime-green lines. The final fit of the spectrum is presented in red. The smaller panel shows the continuum subtracted total fit in red. The broad components contributing to the total fit are presented in blue and the narrow components are presented in green.}
 \label{fig:Fit_Spec_J}
\end{figure}

The $J$-band contains the narrow [\ion{O}{ii}] line. The [\ion{O}{ii}] doublet is spectrally resolved and provides a good indicator of the systemic redshift. The previous best systemic redshift estimate of \ulas was calculated using the CO(3$-$2) emission line from ALMA data, yielding $\rm z_{sys} = 2.566$ \citep{2021MNRAS.503.5583B}. We fit the [\ion{O}{ii}] emission doublet without the PyQSOFit iron template, since the wavelength coverage is too restricted to constrain the fit. A set of broad and a set of narrow components, whose widths and velocity offsets are tied, are used to represent the quasar and galaxy emission. We present the results in Fig.\ref{fig:Fit_Spec_J}. We estimate a redshift, $\rm z_{sys} = 2.5656 \pm 0.0005$ from [\ion{O}{ii}] which is entirely consistent with the value derived from the CO(3-2) line and is adopted as the systemic redshift for the remainder of the paper. 

We can also estimate a value for the instantaneous, unobscured SFR of any host galaxy component using [\ion{O}{ii}] emission line luminosity by assuming that the line flux does not have any contributions from the AGN. We note however that even under this assumption, the [\ion{O}{ii}] line flux can be significantly affected by both dust extinction and metallicity. Since it is unclear to what degree the line is affected by dust extinction relative to the quasar continuum, we estimate lower limits on the [\ion{O}{ii}] line luminosity and SFR([\ion{O}{ii}]), assuming E(B-V)=0 towards the star-forming regions in the host galaxy. We calculate the SFR([\ion{O}{ii}]) with Eqn. \ref{eq:SFR};

\begin{equation}
    \rm SFR_{[O\,\textsc{ii}]}(\emph{M}_{\odot}\, yr^{-1}) = (1.4 \pm 0.4) \times 10^{-41}  L([\ion{O}{ii}]) 
    \label{eq:SFR}
\end{equation}

\noindent where the [\ion{O}{ii}] line luminosity L([\ion{O}{ii}]) = $10^{42.34\pm0.10}$ $\rm erg\;s^{-1}$ for \ulas \citep{1998ARA&A..36..189K}. Eqn. \ref{eq:SFR} then yields a SFR$_{\rm [O\,\textsc{ii}]}$ = $31.1 \pm 9.3$ $ M_{\odot} \rm yr^{-1}$.

\subsubsection{Black-hole mass, Eddington ratio and narrow-line region outflows}

\begin{figure*}
 \includegraphics[scale=0.85]{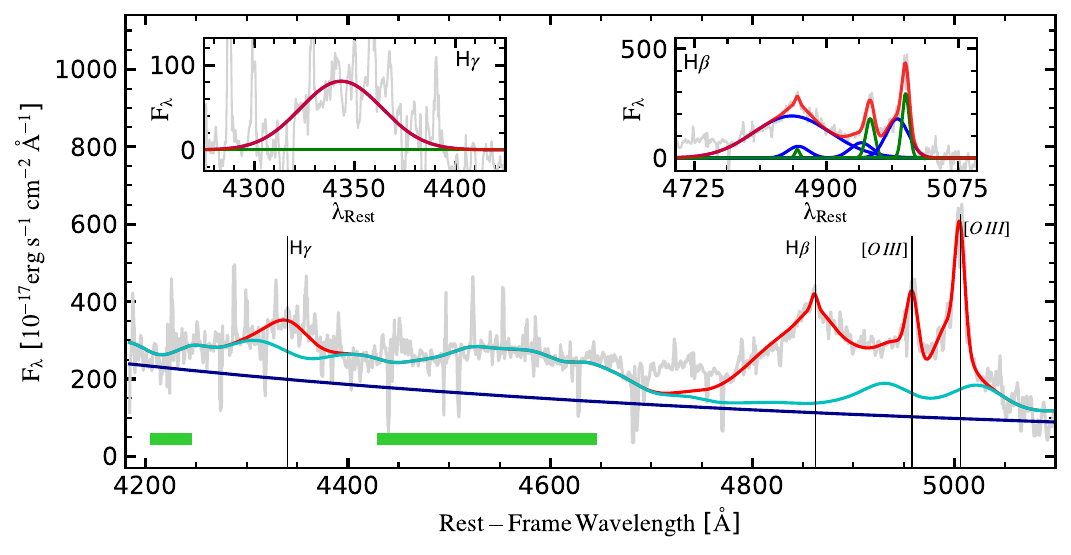} 
  \caption{We present the $H$-Band spectrum for \ulas with a dust correction, E(B-V)$^{\rm QSO}$ = 1.55, applied in grey. The continuum is presented in navy blue, and the windows over which the continuum was fitted are denoted by the thick, lime-green lines. In addition, the combination of the continuum and iron template are presented in teal. The final fit of the spectrum is presented in red. Noteworthy emission lines are marked by thin black lines and labelled appropriately. The smaller panels are labelled by the line/complex they represent. The continuum subtracted total fit is presented in red, the broad components contributing to the total fit are presented in blue and finally, the narrow components are presented in green.}
 \label{fig:Fit_Spec_H}
\end{figure*}

\begin{figure}
 \includegraphics[width=\linewidth]{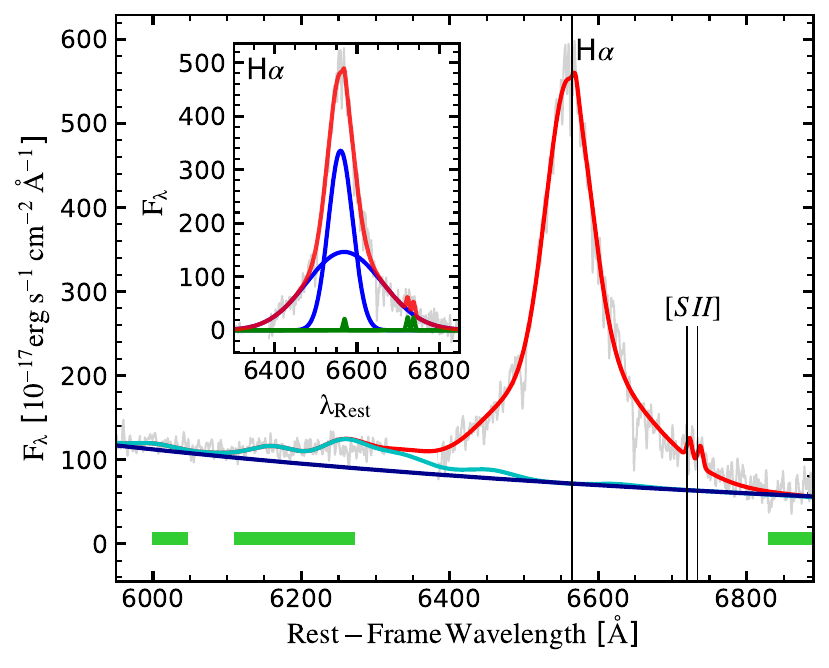} 
  \caption{Same as Fig.\ref{fig:Fit_Spec_H}, but instead we present the $K$-Band spectrum.}
 \label{fig:Fit_Spec_K}
\end{figure}

From Section \ref{SEC:SED_Fit}, it is clear that redward of 4200\AA\ the spectrum is dominated by the dust-reddened Type 1 QSO component. For this reason, we apply a dust correction, using the E(B-V)$^{\rm QSO}$ = 1.55 estimate from Table \ref{tab:SED_Best_fit} to both the $H$ and $K$ bands before fitting the spectral lines. We model the H$\gamma$ emission with a single broad component. The H$\beta$ emission was modelled by two broad components and an additional narrow component. Both [\ion{O}{iii}] lines were modelled by a single broad component paired with an additional narrow line contribution whose widths and velocity offsets were tied. In the $K$-Band, we model H$\alpha$ with two broad components and a single narrow component. Additionally, we model the [\ion{S}{ii}]$\lambda\lambda6716,6730$ emission doublet with a further two narrow components, whose widths and velocity offsets are tied. Due to the extreme H$\alpha$ emission strength, we are unable to robustly model the narrow [\ion{N}{ii}] emission. The results are presented in Fig.\ref{fig:Fit_Spec_H} and Fig.\ref{fig:Fit_Spec_K}.

Using these fits, we estimate the equivalent width (EW), full-width half-maximum (FWHM), emission line blueshifts and signal-to-noise ratio for each emission line. We define emission line blueshifts using Eqn. \ref{eq:blueshift};

\begin{equation}
    \rm V_{50} = c(\lambda_r -\lambda_{\textnormal{50}})/\lambda_r \, \, \,[\textnormal{km\,s}^{-1}]
    \label{eq:blueshift}
\end{equation}

\noindent where c is the velocity of light, $\lambda_{\textnormal{50}}$ is the rest-frame wavelength that bisects the 50$^{\textnormal{th}}$ percentile line flux and $\lambda_r$ is the rest-frame wavelength of a given emission feature. The use of line centroids to define blueshifts reflects the well-known line asymmetries of some emission features such as \ion{C}{iv} \citep{Richards_2011,2020MNRAS.492.4553R, 2023MNRAS.523..646T,2023MNRAS.524.5497S}. Furthermore, for the [\ion{O}{iii}] emission, we also calculate the W80$=$V90$-$V10, since the [\ion{O}{iii}] emission line width is often used as an indicator of narrow-line region outflows \citep[e.g.][]{2016ApJ...817...55S,2019MNRAS.488.4126P,2019MNRAS.486.5335C,2019MNRAS.487.2594T,2020A&A...634A.116V}. V10 and V90 are calculated using Eqn. \ref{eq:blueshift}, evaluated at the 10$^{\textnormal{th}}$ and 90$^{\textnormal{th}}$ percentile of the line flux, respectively. The S/N of the emission lines are estimated by calculating the ratio between the maximum continuum-subtracted flux and the mean noise across the emission line feature. A summary of the quasar emission line properties can be found in Table \ref{tab:Line_prop}.

We observe broad blue wings associated with the [\ion{O}{iii}] emission in Fig.\ref{fig:Fit_Spec_H}, which is reflected in the large line width, W80 = 1830$\pm$350 $\textnormal{km\,s}^{-1}$. The estimated [\ion{O}{iii}] line luminosity, L([\ion{O}{iii}]) = $10^{43.71\pm0.04}$ $\rm erg\;s^{-1}$. Our estimate of L([\ion{O}{iii}]) is sensitive to dust extinction, hence this value serves as a lower limit assuming E(B-V)=0 since we do not know to what extent the narrow-line region is affected by the attenuation suffered by the quasar continuum. The presence of significant line blueshifts in the [\ion{O}{iii}] emission is consistent with the high UV/optical continuum luminosity estimated by the SED fitting in Section \ref{SEC:SED_Fit} and the large L([\ion{O}{iii}]). The [\ion{O}{ii}] emission centroid is located at the systemic redshift by definition.

\begin{table}
    \centering
    \caption{Quasar emission line properties inferred from the \ulas X-Shooter spectrum. Since the [\ion{O}{ii}] emission was used to determine z$_{\rm sys}$, its V50 velocity is zero by definition. Here, the results pertaining to the [\ion{O}{iii}] emission refer to the $\lambda$5008\AA\, line.}
    
    \begin{tabular}{|l|c|c|c|c|c|}
         \hline
         \hline
            Species & S/N & REW [\AA] & FWHM [$\textnormal{km\,s}^{-1}$]  & V50 [$\textnormal{km\,s}^{-1}$]   \\
         \hline
         \hline
            Ly$\alpha$ & 254 & 97$\pm$2 & 1070$\pm$10 & -350$\pm$20  \\
            \ion{N}{v} & 11.9 & 16$\pm$1 & 4227$\pm$3  & -930$\pm$210  \\
            \ion{O}{iv}/\ion{Si}{iv} & 4.3 & 7.0$\pm$0.3 & 5773$\pm$6 & -940$\pm$160  \\
            \ion{C}{iv} & 16.1 & 24$\pm$1 & 3900$\pm$390  & -1080$\pm$110 \\
            \ion{O}{ii} & 12.8  & 8$\pm$2 & 330$\pm$120  & 0  \\
            H$\gamma$ & 2.5 & 21$\pm$1 & 3340$\pm$190 & +60$\pm$20  \\
            H$\beta$ & 23.9 & 230$\pm$30 & 6030$\pm$580  & -120$\pm$50  \\
           \ion{O}{iii} & 37.0 & 68$\pm$12 & 940$\pm$80  & -380$\pm$90  \\
            H$\alpha$ & 42.8 & 820$\pm$10 & 4040$\pm$70  & -320$\pm$30  \\
         \hline
         \hline  
    \end{tabular}
    \label{tab:Line_prop}
\end{table}

The Balmer lines are very broad as seen in Table \ref{tab:Line_prop}. We calculate the SMBH mass, M\textsc{bh}, of \ulas from the measured width of the H$\beta$ emission using the following equation \citep{2006ApJ...641..689V};

\begin{equation}
    \begin{aligned}
     \rm log(H\beta, M\textsc{bh} [M_{\odot}]) = &\, \rm log\left\{\left[\frac{FWHM(H\beta)}{1000\, \textnormal{km\,s}^{-1}}\right]^{2}\left[\frac{\lambda L_\lambda(5100\,\mathring{A})}{10^{44}\, erg\;s^{-1}}\right]^{0.50}\right\} \\
     & + (6.91 \pm 0.02) \\
    \end{aligned}  
    \label{eq:Hb_mass}    
\end{equation}

\noindent where $\rm \lambda L_\lambda(5100\,$\AA) = $ \rm 10^{47.51}\, erg\;s^{-1}$ is the continuum luminosity of \ulas evaluated at 5100\AA,\, using the best-fit continuum from PyQSOFit in the $H$-Band. Eqn. \ref{eq:Hb_mass} yields $\rm log_{10}(H\beta, M\textsc{bh} [M_{\odot}]) = 10.26 \pm 0.05$, consistent with the analysis of a much shorter exposure SINFONI spectrum of this target in \citet{2015MNRAS.447.3368B}. Using the FWHM correction between H$\beta$ and H$\alpha$ detailed in \citet{2005ApJ...630..122G}, we can also estimate M\textsc{bh} from the H$\alpha$ emission, yielding $\rm log_{10}(H\alpha, M\textsc{bh} [M_{\odot}]) = 10.17 \pm 0.09$. The uncertainties in $\rm \lambda L_\lambda(5100\,$\AA) were calculated by generating 500 simulated spectra perturbed with Gaussian noise consistent with the noise spectrum and implementing PyQSOFit after each iteration. Uncertainty estimates in M\textsc{bh} were then calculated by propagating the uncertainty in the continuum luminosity through Eqn.\ref{eq:Hb_mass}, drawing the normalisation constant from a Gaussian distribution with $\rm \mu = 6.91\, \&\, \sigma = 0.02$. This does, however, neglect the large ($\sim$0.5 dex) systematic uncertainties reported in single-epoch M\textsc{bh} estimates by \citet{2006ApJ...641..689V}. Previous estimates of the dynamical mass of this system based on CO(3-2) observations suggest $\rm log_{10}(M_{dyn}/M_{\odot}) \simeq 11$ \citep{2021MNRAS.503.5583B}. The SMBH in \ulas is therefore over-massive with respect to the host galaxy but broadly consistent with the scaling relation for high-redshift luminous quasars \citep[e.g.][]{2020A&A...637A..84P}.

Using the relation between $\rm L_\lambda(5100\,$\AA) and $\rm L_{Bol}$ from \citet{2019MNRAS.488.5185N} we define a bolometric correction $\rm BC_{5100}$ = 4.46, yielding $\rm L_{Bol}$ = $ \rm 10^{48.16}\, erg\;s^{-1}$ for \ulas. Using this bolometric luminosity and our estimate of the black-hole mass yields an Eddington-scaled accretion rate of log$_{10}$($\rm \lambda_{Edd}) = -0.19$. Given the intrinsic X-ray luminosity $\rm L_{X,int}$ = $ \rm 10^{45.6}\, erg\;s^{-1}$ for \ulas \citep{2020MNRAS.495.2652L}, the corresponding X-ray bolometric correction $\rm log_{10}(k_{Bol},L_X) = 2.56$. While this is extreme, X-ray bolometric corrections as large as $\rm k_{Bol,X} = 100-1000$ have already been observed in the WISE/SDSS selected hyper-luminous (WISSH) quasar sample with bolometric luminosities $\rm L_{Bol} > 2 \times 10^{47}\, erg\;s^{-1}$. Indeed, the inferred X-ray bolometric correction is consistent with the trends in bolometric correction with both black-hole mass and Eddington-scaled accretion rate inferred in other hyper-luminous infrared-selected quasar populations \citep{2017A&A...608A..51M}. Similarly, the mid-infrared luminosity inferred from Wide-field Infrared Survey Explorer (WISE) photometry, $\rm log_{10}(L_{6\mu m}) = 47.5$, suggests a mid-infrared bolometric correction $\rm log_{10}(k_{Bol},L_{6\mu m}) = -0.66$. This is consistent with the $\rm log_{10}(k_{Bol},L_{7.8\mu m}) = -0.35 \pm 0.4$ reported amongst blue AGN \citep{2012ApJ...761..184W}.

\subsubsection{UV Emission and Absorption Line Features}

\begin{figure*}
 \includegraphics[scale=0.85]{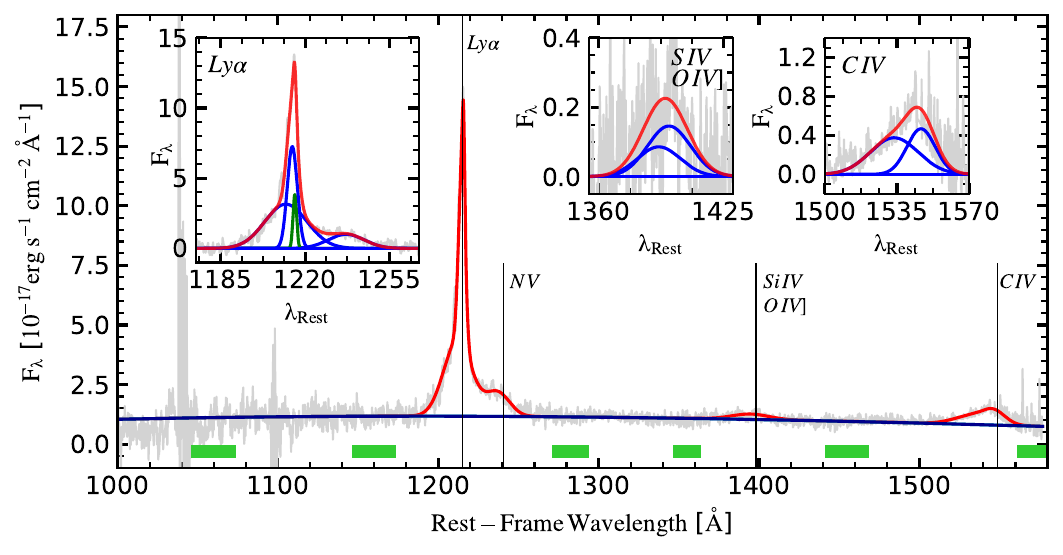} 
  \caption{Same as Fig.\ref{fig:Fit_Spec_H}, but instead we fit the UVB  spectrum without a dust correction applied.}
 \label{fig:Fit_Spec_UVB}
\end{figure*}

In Section \ref{SEC:SED_Fit} we concluded that the broad emission lines seen in the UVB can be explained by a scattered or leaked quasar component. Hence we do not apply any dust correction to the UVB spectrum before fitting the lines using PyQSOFit. We model three main broad-line complexes in the UVB. We parametrise the Ly$\alpha$/\ion{N}{v} complex with four Gaussians, one broad component attributed to the \ion{N}{v} contribution and the remainder attributed to the Ly$\alpha$ contribution. The \ion{Si}{iv}/\ion{O}{iv]} complex is modelled by two broad components and so too is the \ion{C}{iv} emission. To enable a more robust reconstruction of the emission line profile, narrow absorption features must first be masked. We adopt an iterative approach for this. First we fit the raw spectrum with PyQSOFit, the result of which serves as a pseudo-continuum for subsequent fits. Pixels whose flux $>2\sigma$ below the pseudo-continuum are replaced by the corresponding flux element in the pseudo-continuum itself. The process is then repeated until successive models converge (in this case, five iterations were sufficient). The results are presented in Fig.\ref{fig:Fit_Spec_UVB} and the emission line properties are summarised in Table \ref{tab:Line_prop}.

In Fig.\ref{fig:Fit_Spec_UVB} the \ion{C}{iv} emission line appears broad, blueshifted and of modest EW. These \ion{C}{iv} properties are typical of those found in luminous, blue, Type 1 QSOs and have been associated with broad line region (BLR) winds along certain sight-lines \citep{Richards_2011, 2020MNRAS.492.4553R, 2023MNRAS.523..646T,2023MNRAS.524.5497S}. The \ion{N}{v} emission is also blueshifted. As \ion{N}{v} has a similar ionising potential to \ion{C}{iv}, the two species are likely emitting from similar regions in the BLR, hence, the broadly consistent line properties are to be expected. Conversely, the Ly$\alpha$ EW is uncharacteristically large in comparison to \ion{N}{v}. The velocity off-set of the narrow component of the Ly$\alpha$ emission (green Gaussian in top left panel of Fig. \ref{fig:Fit_Spec_UVB}) is --16$\pm$170 $\rm km\,s^{-1}$, consistent with the uncertainties on the systemic redshift. We measure a line ratio \ion{N}{v}/Ly$\alpha\,\sim 0.16$ for \ulas, significantly less than the \ion{N}{v}/Ly$\alpha\,\sim 0.31$ observed in blue Type 1 QSOs \citep{2016ApJ...817...55S}. We propose that the line profile can be explained either by a contribution from a star-forming galaxy (as discussed in Section \ref{SEC:SED_Fit}) or some scattering of the quasar light. For a full discussion see Section \ref{SEC:DISC_UV_CONT}.

We searched for absorption features associated with a number of high ionisation species in the UV spectrum as follows. First the spectrum is renormalised by a pseudo-continuum defined by the best PyQSOFit broad emission line model (Fig.\ref{fig:Fit_Spec_UVB}, red). We then employ a cross-correlation technique similar to that described in \citet{2010yCat..74052302H, 2020MNRAS.492.4553R,2023MNRAS.524.5497S}. Our default model is defined as F/F$_{\rm cont}=1$. We then define a grid of Gaussian absorption line models with various line depths and widths (the narrowest model explored has width $\sigma \rm \sim30\,km\,s^{-1}$ - equal to the resolution of X-Shooter in the UVB arm). For doublets, the line widths and line off-sets are tied and we additionally explore the following line ratios; 1:3, 1:2, 2:3, 1:1. The best model parameters and corresponding cross-correlation value are recorded at each line off-set. The absorption line model whose combination of off-set, depth, width and line ratio yields the maximum cross-correlation value is added to the default model and the process is then repeated. If the best absorption model does not represent a >$4\sigma$ improvement in cross-correlation value over the default model or the new absorption feature does not exceed a signal-to-noise of three (S/N$\,>3$), the new absorption line model is rejected and the search is complete. 

\begin{figure*}
 \includegraphics[scale=0.6,trim={0 1.0cm 0 0}]{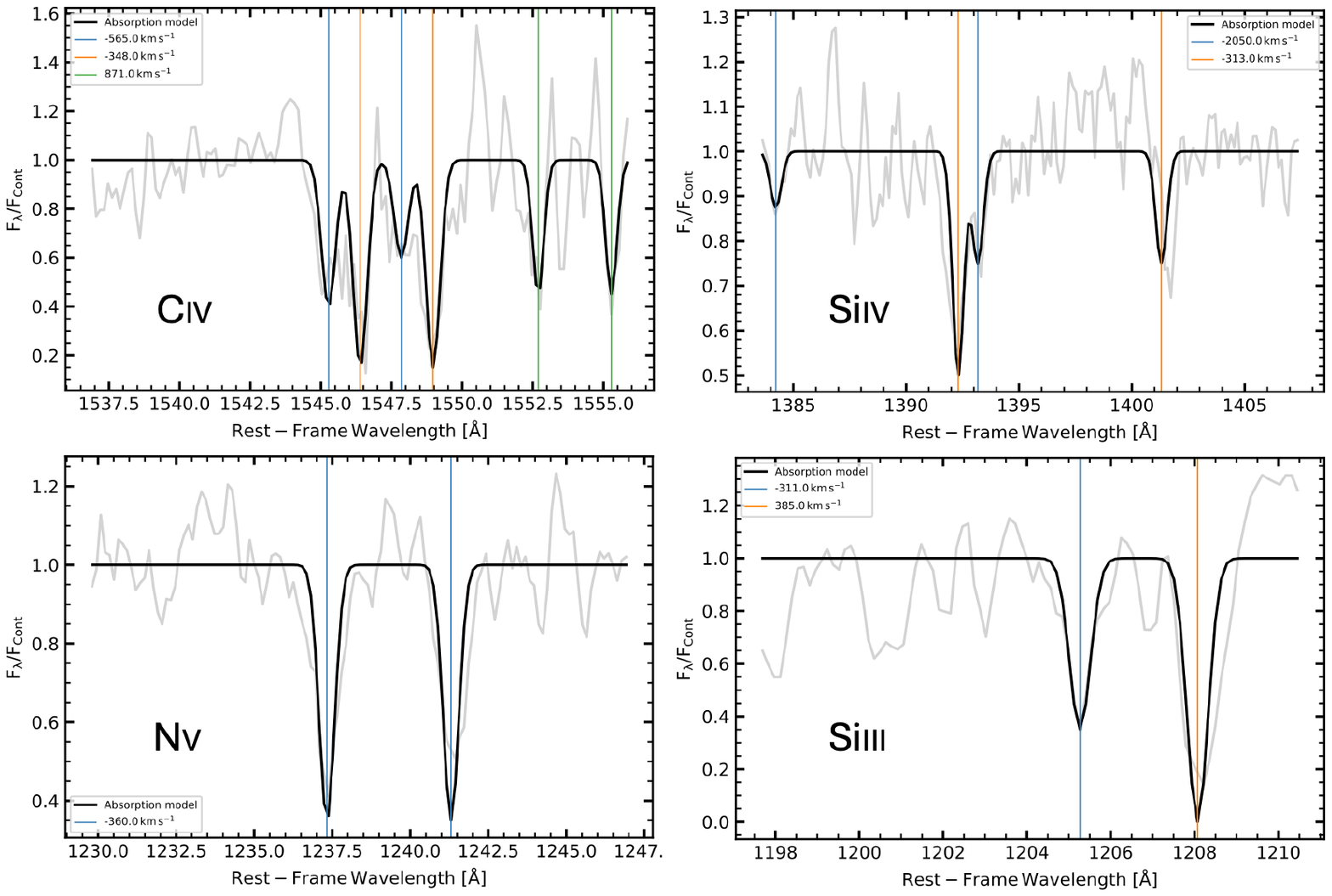} 
  \caption{The UVB narrow absorption line features detected in the X-Shooter spectrum. The flux density normalised by the PyQSOFit pseudo-continuum is presented in grey. The absorption model reconstructions are presented in black. The species are labelled in each panel. We mark the various velocity components with vertical lines in each panel. }
 \label{fig:UVB_absorption}
\end{figure*}

The results are summarised in Table \ref{tab:Absorption} and Fig. \ref{fig:UVB_absorption}. Five distinct velocity components are detected, with component 3 at $v\simeq-340 \rm km\,s^{-1}$ detected in all species, suggestive of outflowing gas at a range of ionisation parameters. The absorption in \ion{C}{iv} at this velocity is saturated, suggesting a high covering fraction of \ion{C}{iv} gas. In other species, such as \ion{N}{v} and \ion{Si}{iv}, the covering fraction of the gas is much lower. We also see additional absorption components in \ion{C}{iv} at velocities of $\simeq$-565 $\rm km\,s^{-1}$ and $\simeq$+871 $\rm km\,s^{-1}$ that are not present in \ion{N}{v} or \ion{Si}{iii}$\lambda$1206.52 and \ion{Si}{iv}, and have a lower covering fraction relative to the -340 $\rm km\,s^{-1}$ component. Furthermore, we observe a very high velocity component in the \ion{Si}{iv} absorption, with a velocity $\simeq$-2050 $\rm km\,s^{-1}$. We also note that the redshifted  \ion{Si}{iii}$\lambda$1206.52 component at +385 $\rm km\,s^{-1}$ could conceivably also represent blueshifted Ly$\alpha$ absorption.

\begin{table}
    \centering
    \caption{Summary of different velocity components of narrow associated absorption line systems (AALs) shown in Fig.\ref{fig:UVB_absorption}. Velocity offsets are relative to the [\ion{O}{ii}] systemic redshift. We group velocity components whose off-sets are consistent within the uncertainty on $\rm z_{sys}$.}
    
    \begin{tabular}{|l|c|c|}
         \hline
         \hline
          Species & Component & Offset [$\textnormal{km\,s}^{-1}$] \\
         \hline
         \hline
            \ion{C}{iv} $\lambda\lambda1548.19,1550.77$ & 2 & -565  \\
            & 3 & -348\\
            & 5 & 871 \\
        \hline
            \ion{Si}{iv} $\lambda\lambda1393.76,1402.77$ & 1 & -2050  \\
            & 3 & -313\\
        \hline
            \ion{N}{v} $\lambda\lambda1238.82,1242.80$ & 3 & -360 \\
        \hline
            \ion{Si}{iii} $\lambda1206.52$ & 3 & -311  \\
            & 4$^a$ & 385 \\
         \hline
         \hline
         \multicolumn{3}{l}{$^a$ Potentially blueshifted \ion{H}{i} absorption, $\rm V_{off-set}$ = -1870$\textnormal{km\,s}^{-1}$}
    \end{tabular}
    \label{tab:Absorption}
\end{table}

\section{Discussion} \label{SEC:DISC} 

\subsection{Source of the UV emission} \label{SEC:DISC_UV_CONT} 

In Section \ref{SEC:SED_Fit}, we concluded that a simple dust-attenuated Type 1 QSO SED model is unable to reproduce the emission blueward of $\sim 4000$\AA\, in \ulas and that emission from a secondary blue AGN component and/or a star-forming host galaxy likely contributes in this region. We now further explore the origin of the excess UV emission. 

\subsubsection{A dual AGN system?} 

Could the blue AGN component seen in the X-Shooter spectrum of \ulas be consistent with the presence of a second AGN in the system, which is more than 1000 times fainter in the UV than the primary dusty quasar? \ulas is hosted in a major merger based on the ALMA observations reported in \citet{2021MNRAS.503.5583B}. The two galaxies in the merger are both detected in CO(3-2) and are separated by $\sim$ 15 kpc in projection. During galaxy mergers, when both of the SMBHs are activated, a dual quasar could be formed \citep{begelman1980massive}. Observations over the past decades have reported hundreds of dual quasars with redshifts from local to z=5.66 \citep{yue2021candidate}, and separations from Mpc scale down to 230 pc (\citet{koss2023ugc}. Therefore a secondary quasar could be a plausible source of the UV emission.

If a secondary AGN is present however, it is likely not associated with the companion merging galaxy detected in CO with ALMA. The merging companion galaxy has gas excitation conditions that are more typical of star-forming galaxies rather than AGN as well as narrower molecular emission lines compared to the quasar host \citep{2018MNRAS.479.1154B}, all of which seems to suggest no actively accreting black hole in this galaxy. Moreover, the image decomposition presented in Fig. \ref{fig:galight} suggests the presence of at most a single point-source contributing to the UV emission, located at the centroid of the primary quasar host rather than the merging companion. If this is a dual AGN system, then the separation between the secondary AGN and the quasar host galaxy must therefore be less than the typical seeing of the HSC images (0.6 arcsec, or 4.8 kpc at z=2.566). The observational constraints of dual fractions at this separation are still poor, due to the limited resolution \citep[e.g.,][]{silverman2020dual}. On the other hand, \cite{steinborn2016origin} studied the statistics and properties of closely-separated (<10 kpc) dual AGN at z=2 in the Magneticum simulation. This cosmological simulation encloses a volume of 182 Mpc$^3$ and produces 35 BH pairs, among which nine are dual AGN. Thus the spatial density of dual AGN is $\sim$0.05 Mpc$^{-3}$. They also report that the dual fraction (the ratio of the number of dual AGN and the total number of AGN) is $1.2\pm0.3\%$ in the simulation \citep[also see][this fraction varies roughly between 0.1-5\%]{volonteri2016cosmic,rosas2019abundances}. Therefore, the dual AGN scenario, although cannot be fully ruled out, is statistically unlikely.

\subsubsection{Leaked or scattered light from the primary quasar?}

Since the AGN emission is unlikely to originate from a secondary source, the broad UV emission most likely originates from the primary quasar. Here we will discuss two potential scenarios - (i) leaked AGN emission, escaping through "windows" in a patchy dust morphology and (ii) scattered AGN emission.

In Section \ref{SEC:SED_Fit}, we show that when modelling the rest-frame UV portion of the SED with an unattenuated AGN component, just 0.05 per cent of the intrinsic quasar emission is required to reproduce the continuum. Since the UV continuum is consistent with the shape of a blue AGN, we require a consistent covering factor across all wavelengths to preserve the shape of the the accretion disk's multi-temperature black body in a "leaked" light scenario. For such a small fraction of the intrinsic radiation to be observed, we would require the obscuring medium to populate the extremely local regions of the AGN. The required dust morphology is then extremely contrived and hence we conclude that leaked AGN emission through a patchy obscuring medium is an unlikely cause of the UV-excess \citep{2015ApJ...804...27A}. 

An alternative scenario is that AGN emission is scattered towards the line-of-sight by the obscuring medium. This is the favoured explanation in HotDOGs, where broad-band UV photometry confirms that the emission is linearly polarised, having been scattered from ionised gas \citep{2015ApJ...804...27A,2020ApJ...897..112A,2022ApJ...934..101A}. However, unlike HotDOGs and ERQs, \ulas shows evidence of high-ionisation line blueshifts in the rest-frame UV emission (See Fig.\ref{fig:Fit_Spec_UVB} and Table \ref{tab:Line_prop}). While this does not completely rule out the scattered light scenario, it does limit the number of feasible dust geometries. 

\begin{figure}
 \includegraphics[width=\linewidth]{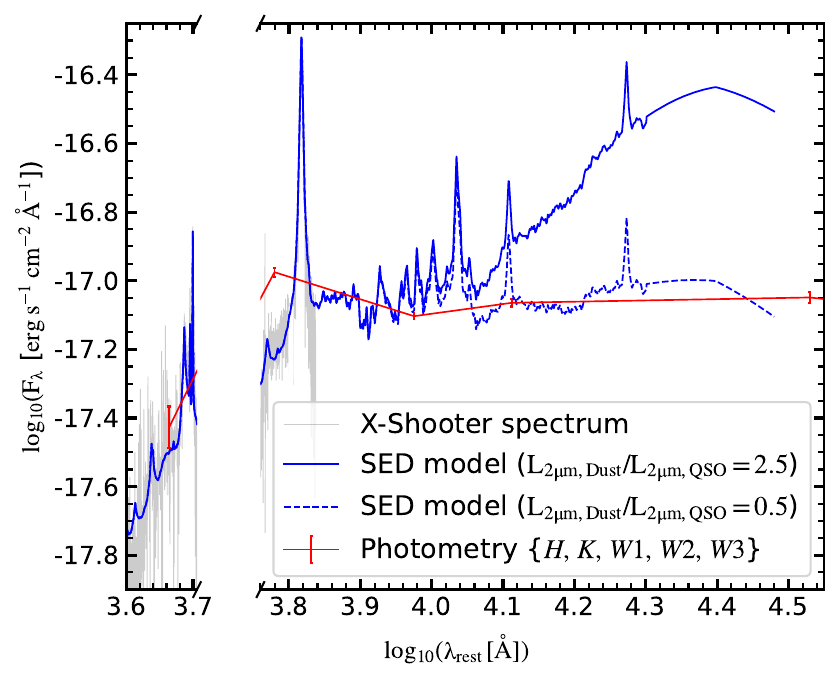} 
  \caption{The X-Shooter spectrum of \ulas is presented in grey, with the broad-band photometry from UKIDSS and WISE overlaid in red. The best-fit UV/optical SED, see Section \ref{SEC:SED_Fit}, is extrapolated to mid-infrared wavelengths (blue, solid), assuming $\rm L_{2\mu m, Dust}/L_{2\mu m, QSO}=2.5$. The photometry is inconsistent with the SED model redward of the $W1$ filter. The same SED model assuming $\rm L_{2\mu m, Dust}/L_{2\mu m, QSO}=0.5$ (blue, dashed) is more consistent with the photometry. Both SED models assume a hot dust temperature of 1243K. This suggests that the \ulas system contains less hot dust than the typical blue SDSS quasar, where $\rm L_{2\mu m, Dust}/L_{2\mu m, QSO}\gtrsim1.0$ \citep{2021MNRAS.501.3061T}.}
 \label{fig:MIR_Phot}
\end{figure}

Should the obscuring medium be distant enough such that the incident AGN emission features only the sight lines with blue-shifted high-ionisation lines, the scattered spectrum could also exhibit blueshifts. A candidate for the obscuring medium is then some kind of dusty toroidal structure outside the BLR. In  Fig.\ref{fig:MIR_Phot} we show the infrared photometry of \ulas from UKIDSS and \textit{WISE} compared to the best-fit SED from Section \ref{SEC:SED_Fit}. The infrared emission in the SED model represents the average infrared SED for an SDSS quasar at $z\sim2$ \citep{2021MNRAS.501.3061T}
with a hot dust temperature of T$_{bb}$ = 1243K and a hot dust normalisation - defined as the ratio between the hot dust and accretion disk luminosities at 2$\,\mu$m - $\rm L_{2\mu m, Dust}/L_{2\mu m, QSO}=2.5$. We can immediately see that this SED model is inconsistent with the photometry of \ulas when extrapolated to mid-infrared wavelengths. A value of $\rm L_{2\mu m, Dust}/L_{2\mu m, QSO}=0.5$ however is better able to match the observed photometry. The flux density observed in the $W4$-Band (not shown in Fig.\ref{fig:MIR_Phot}) is $10^{-17.1\pm0.1} \rm erg\,s^{-1}\,cm^{-2}\,\mathring{A}^{-1}$. The mid-infrared SED therefore remains flat to a rest-frame wavelength $\sim 6\mu m$ (corresponding to a dust temperature of $\rm \sim 450K$). We therefore conclude that the ratio between the hot dust and accretion disk luminosities in \ulas is significantly lower than observed in blue quasars \citep[see][]{2021MNRAS.501.3061T}. 

\begin{figure*}
 \includegraphics[width=\linewidth, trim={0 3.5cm 0 0}]{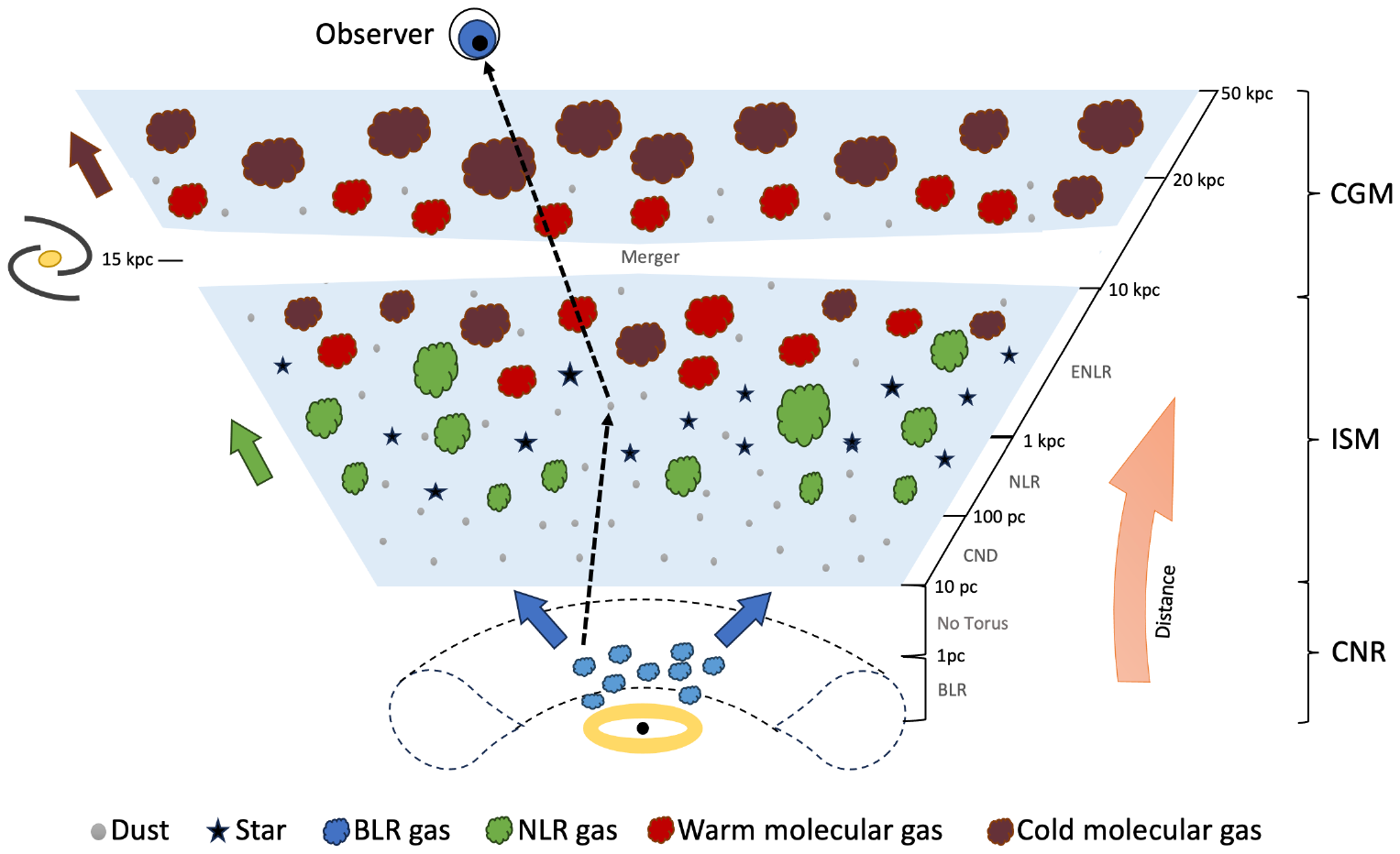} 
  \caption{An illustration of our proposed geometry that best describes \ulas focusing on the different scales associated with different features in the X-Shooter spectrum and the multi-wavelength observations of this source. On sub-pc scales we illustrate the BLR gas with blue clouds. Since the mid-infrared SED is devoid of hot dust (Fig.\ref{fig:MIR_Phot}), we illustrate the missing/depleted torus with dotted black lines stretching to $\sim$10 pc. ALMA observations suggest that the ISM dust is on scales of $\sim$15-20 kpc based on the size of the millimetre continuum emission. NLR gas and star-forming regions are represented with green clouds and black stars respectively. The warm molecular gas is depicted with red clouds and the cold molecular gas is depicted with maroon clouds. A cartoon sight line, scattered from ISM dust and then subsequently absorbed by the warm and cold gas reservoirs, potentially giving rise to the AAL features in the UV spectrum, is illustrated by the dashed black arrow. The merging companion galaxy resides at 15 kpc and is illustrated to the left of the sketch. Blue, green and maroon arrows represent BLR, NLR and cold molecular gas flows respectively all of which exhibit significant velocity offsets relative to the systemic redshift.}
 \label{fig:geometry}
\end{figure*}

The absence of hot dust emission on typical torus scales could suggest that the obscuring medium responsible for the significant (E(B-V)$^{\rm QSO}$ = 1.552 mag) extinction towards the quasar continuum is likely on ISM, rather than nuclear scales. Alternatively, self-absorbed cold torus models have been invoked to explain flatter mid-infrared SEDs \citep[e.g.][]{2019ApJ...884..171H}. However we might expect a more symmetrical line profile in \ion{C}{iv}, with weaker blueshifts should the scattering medium lie primarily on nuclear scales close to the quasar BLR. Hence, we favour the depleted torus scenario, proposing a plausible geometry for \ulas in Fig.\ref{fig:geometry}. A more complete analysis of the mid infra-red SEDs of the HRQ population is deferred to future work. In Fig.\ref{fig:geometry}, the excess UV emission seen in the spectrum is traced by dashed black arrows. We show multiple sight-lines originating from the accretion disk scattered by interstellar dust towards the observer.

\subsubsection{Associated Absorption Line Systems in \ulas}

In Fig. \ref{fig:UVB_absorption} we can see numerous narrow absorption features in the rest-frame UV spectrum with linewidths of a few hundred $\rm km\,s^{-1}$ and that lie within 3000 $\rm km\,s^{-1}$ of the quasar systemic redshift. These properties are consistent with the associated absorption line (AAL) systems that have been detected in the spectra of numerous luminous quasars (e.g. \citealt{VandenBerk:08, Hamann:11, Shen:12, Chen:18}). AALs can have a wide range of origins tracing gas in quasar inflows and outflows, in the halos of the quasar host galaxy or indeed gas reservoirs on circumgalactic scales \citep{Foltz:86, Tripp:98}. Statistical studies suggest that most AALs are intrinsic to the quasars, and hence can serve as important probes of the quasar environment \citep{Nestor:08, Wild:08}. As such, the detection of significant absorption in this gas-rich merger is perhaps not surprising. Dust-reddened quasars in the Sloan Digital Sky Survey (SDSS) have been associated with a higher incidence of AALs \citep{Richards:03,VandenBerk:08} and 2MASS red quasars with a higher incidence of Broad Absorption Line (BAL) features (e.g. \citealt{2009ApJ...698.1095U}). However, to our knowledge, our study represents the first detection of multiple AAL systems in a quasar with such extreme dust reddening of E(B-V)$>1.5$. 

The most prevalent absorption feature detected in multiple species is at a velocity of $\sim$-340 $\rm km\,s^{-1}$ relative to systemic. The higher ionisation gas at -340 $\rm km\,s^{-1}$ could conceivably be associated with the merging companion, which is located 15 kpc away in projected distance from the quasar and blueshifted in CO(3-2) emission by 170 $\rm km\,s^{-1}$ relative to the quasar \citep{2021MNRAS.503.5583B}. However, the high ionisation gas is most likely located closer to the ionising source, effectively shielding the lower ionisation gas, which is likely spatially co-incident with the warm molecular gas traced by CO(3-2). We also detect redshifted AALs at a velocities of $\simeq$385 $\rm km\,s^{-1}$ and $\simeq$870 $\rm km\,s^{-1}$. These components may be associated with the CO(1-0) cold gas reservoir in this quasar, which is off-set in velocity by $\simeq$500 $\rm km\,s^{-1}$ \citep{2018MNRAS.479.1154B}. The cold gas reservoir is also very spatially extended on projected scales of $\sim$ 50 kpc. The geometry of the system showing the different spatial scales for the gas potentially producing the AAL features is shown in Fig. \ref{fig:geometry}.  

\subsubsection{Star-forming host} 

As discussed in Section \ref{SEC:SED_Fit}, HSC imaging is suggestive of a host galaxy contribution to the rest-frame UV flux. This is also consistent with the strong nebular emission observed in Ly$\alpha$ and [\ion{O}{ii}] as well as the presence of multiple narrow absorption line systems in Fig. \ref{fig:UVB_absorption} which could conceivably be associated with the gas in the quasar host galaxy. The $\rm SFR_{FUV} \sim 3 \times SFR_{\rm [O\,\textsc{ii}]}$. The unobscured SFR, predicted by the far-infrared and submillimeter continuum emission, for this source is $ \rm SFR_{FIR}=680\pm100\, \emph{M}_{\odot}\, yr^{-1}$ \citep{2018MNRAS.479.1154B}. Main sequence galaxies at similar redshifts and A$_{\rm V}^{\rm Gal}$, report similar $\rm SFR_{FIR}$ to \ulas, with their average $\rm SFR_{H\alpha,obs} = 20$ $ M_{\odot}\, \rm yr^{-1}$ consistent with our own estimate of the SFR from [\ion{O}{ii}] \citep{2017ApJ...838L..18P}. Given $\rm SFR_{FUV}$, $\rm SFR_{\rm [O\,\textsc{ii}]}$ and $\rm SFR_{H\alpha,obs}$ estimates make similar assumptions regarding the dust extinction, it would be reasonable to conclude that the [\ion{O}{ii}] estimate is less than than the FUV estimate for \ulas because the rest-frame UV continuum emission in this source does not originate purely from a star-forming host galaxy but rather has some contribution from scattered AGN light as discussed above. 

\subsection{Multi-phase winds in \ulas} \label{SEC:winds}

In this section, we discuss the line properties of \ulas in the context of AGN-driven winds as well as drawing comparisons to observations from other QSO populations. 

There exists a relationship between BLR outflows, as probed by the \ion{C}{iv} blueshift, and both M\textsc{bh} and Eddington-scaled accretion rate in blue quasars. This relationship is evident in both simulations \citep[e.g.][]{2019A&A...630A..94G} and observations of blue Type 1 SDSS quasars up to z$_{\rm sys} \sim 4$ \citep[e.g.][]{2023MNRAS.523..646T,2023MNRAS.524.5497S}. The dependence of \ion{C}{iv} blueshift on the M\textsc{bh}-$\rm \lambda_{Edd}$ plane tells us how BLR outflows are linked to the accretion physics of QSOs. For significant ($\rm >1000\,km\,s^{-1}$) \ion{C}{iv} blueshifts, quasars require $\rm \lambda_{Edd} > 0.2$ and M\textsc{bh} $> 10^9 \rm M_\odot$. Our observations of \ulas are consistent with this picture, with a relatively strong \ion{C}{iv} blueshift, V50 = 1075$\rm km\,s^{-1}$, owing to the large SMBH mass, M\textsc{bh} = $\rm 10^{10.26} M_\odot$, and Eddington-scaled accretion rate, $log_{10}(\lambda_{Edd})=-0.19$. Fundamentally, this suggests that the dependence of BLR outflow velocities on black hole mass and accretion rate is broadly similar in blue and heavily reddened quasars. 

\begin{figure}
 \includegraphics[width=\linewidth]{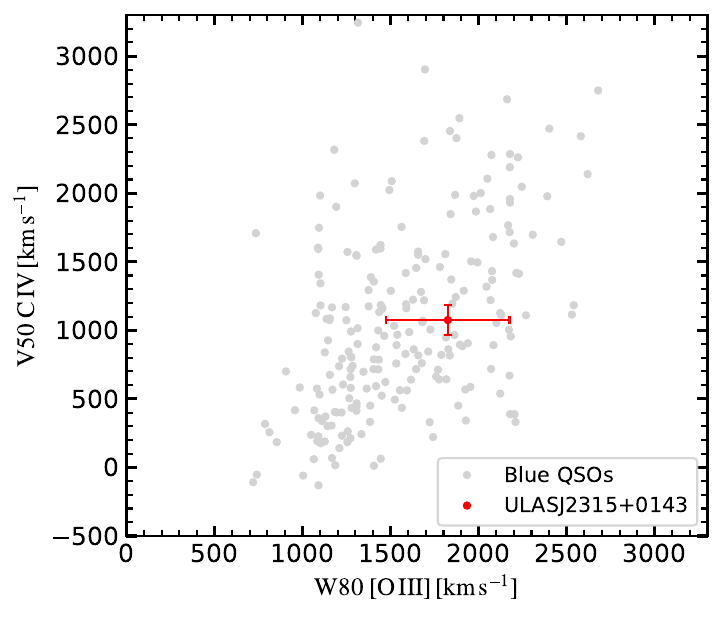} 
  \caption{The \ion{C}{iv} emission line centroid [\ion{O}{iii}] W80 relation for a sample of blue SDSS quasars \citet{2024MNRAS.532..424T} are presented in grey. \ulas, presented in red, shows evidence of significant (though not extreme) BLR and NLR winds, consistent with the trend observed in blue SDSS quasars.}
 \label{fig:W80vsCIV}
\end{figure}

Moving to kpc-scales, we can probe NLR winds with the [\ion{O}{iii}] emission line W80 \citep[e.g.][]{2019MNRAS.488.4126P,2019MNRAS.486.5335C,2019MNRAS.487.2594T,2020A&A...634A.116V}. In Fig.\ref{fig:W80vsCIV}, we compare the BLR and NLR wind velocities of \ulas to a sample of blue QSOs \citet{2024MNRAS.532..424T} and show that \ulas appears consistent with the trend observed in optically selected quasars. While \ulas shows evidence of significant NLR winds, as probed by [\ion{O}{iii}] W80, the object is not extreme. This picture is consistent with the BLR winds as probed by \ion{C}{iv} blueshift. Since the winds in both the broad and narrow line regions are similarly moderate, a common mechanism could be responsible for the gas velocities at these different scales.

In addition, we can harness the [\ion{O}{iii}] emission to investigate the energetics of the outflowing ionised gas. By assuming a symmetric biconical geometry and that the emitting clouds have the same density, \citet{2012A&A...537L...8C} derive the following expression for the mass-loss rate of the NLR outflow;

\begin{equation}
    \begin{aligned}
     \dot M_{out}^{ion} = 164\,\frac{L_{44}([\ion{O}{iii}])\,v_3}{\langle n_{e3}\rangle \,10^{[O/H]}\,R_{\rm kpc}} \,M_\odot\rm\, yr^{-1}\\
    \end{aligned}  
    \label{eq:OIII_Mdot}    
\end{equation}

\noindent where $L_{44}([\ion{O}{iii}])$ is the broad [\ion{O}{iii}] line luminosity in units $\rm 10^{44}\, erg\,s^{-1}$, $v_3$ is the outflow velocity in units of 1000 $\rm km\,s^{-1}$, $\langle n_{e3} \rangle$ is the NLR outflow electron density in units of 1000 $\rm cm^{-3}$, $10^{[O/H]}$ is the oxygen abundance in solar units and R$_{\rm kpc}$ is the radius of the outflowing region in units of kpc. Given the uncertainty in the amount of dust attenuation affecting the NLR, we use the measured [\ion{O}{iii}] luminosity from the spectrum before any dust correction is applied. The kinetic power associated with the outflow is then given by Eqn. \ref{eq:OIII_Edot};

\begin{equation}
    \begin{aligned}
     \dot \epsilon_{k}^{ion} = 5.17\times10^{43}\,\frac{C\,L_{44}([\ion{O}{iii}])\,v_3^3}{\langle n_{e3}\rangle \,10^{[O/H]}\,R_{\rm kpc}} \,\rm erg\,s^{-1}\\
    \end{aligned}  
    \label{eq:OIII_Edot}    
\end{equation}

\noindent where $C = \langle n_{e3}\rangle ^2/\langle n_{e3}^2\rangle \approx 1$. As in \citet{2019MNRAS.488.4126P}, we assume that since the [\ion{O}{iii}] emission is not core dominated in \ulas, $v_{98} = 2521\,\rm km\,s^{-1}$ serves as a good indicator for the outflow velocity and adopt solar metallicities. Using the line ratios between the narrow and broad components of [\ion{O}{iii}] (Fig.\ref{fig:Fit_Spec_H}), we attribute $\sim75$ per cent of the total line luminosity to the broad outflow component. \citet{2017A&A...598A.122B} find that amongst the WISSH broad-line quasar sample $\langle n_{e3} \rangle \sim200 \rm cm^{-3}$ and the [\ion{O}{iii}] emission regions are extended to 1-7 kpc. Since we have no spatially resolved data for the [\ion{O}{iii}] emission in \ulas at present, we adopt the 4 kpc threshold for strong [\ion{O}{iii}] winds reported in another HRQ from our sample in \citet{2012MNRAS.427.2275B} - ULASJ1002+0137 - which was observed by \citet{2023ApJ...953...56V}. Applying these assumptions to Eqn. \ref{eq:OIII_Mdot} and Eqn. \ref{eq:OIII_Edot} yields $\dot M_{ion}^{out} = 199 \,M_\odot\, \rm yr^{-1}$ and $\dot \epsilon_{k}^{ion} = 10^{44.61} \rm erg\,s^{-1} \sim 0.001\,L_{Bol}$. 

Our estimate for the kinetic power of the ionised gas outflows is consistent with the blue quasars presented in \citet{Shen_2011}, given the bolometric luminosity of \ulas, however, they seems relatively weak when compared to the ERQs presented in \citet{2019MNRAS.488.4126P}. It is important though to be mindful of the different assumptions made in the kinetic power estimates of the two samples. Firstly in ERQs the entire [\ion{O}{iii}] line luminosity is attributed to the outflow whereas we attribute $\sim75$ per cent of the total line luminosity to the broad outflow component based on our line fits. Secondly, \citet{2019MNRAS.488.4126P} adopt R$_{\rm kpc} = 1$ since two of their ERQ's [\ion{O}{iii}] emission regions are spatially unresolved down to a resolution of $\sim 1.2$kpcs by the W.M. Keck Observatory OSIRIS integral field spectrograph (IFS) significantly smaller than our assumption of 4 kpc. Both of these assumptions are likely to overestimate the kinetic power of the outflows by a factor of several relative to our assumptions. In addition, if we assume the colour [$i-W3_{\rm Vega}$] is a good indicator of the dust extinction in these systems ([$i-W3_{\rm Vega}]\sim7.78$ for \ulas ), \ulas suffers more extinction than the average ERQ. Hence the [\ion{O}{iii}] line luminosity may be more attenuated in \ulas, resulting in an underestimate of the kinetic power when compared to the ERQ sample. Neither estimate corrects the observed [\ion{O}{iii}] luminosity for dust extinction. Given these assumptions, the ionised outflows observed in \ulas are likely similar to those observed in ERQs in terms of mass-loss rate and kinetic power.

\subsection{Comparison to JWST Little Red Dots}

Recent \textit{JWST} observations point to a significant population of \emph{"little red dots"} (LRDs) at $\rm z_{sys} \gtrsim 5$ with blue rest-frame UV photometric colours - e.g. \citet{Onoue_2023,Kocevski_2023,2024ApJ...964...39G,2024arXiv240403576K}. While the true nature of some sub-samples of LRDs is still debated \citep[e.g.][]{2024ApJ...963..129M}, 60 per cent of the objects from the UNCOVER field show definitive evidence of broad (FWHM > 2000 $\rm km\,s^{-1}$) H$\alpha$ emission and are hence classified as dust-reddened AGN \citep{2024ApJ...964...39G}. Here, we consider whether \ulas is a cosmic-noon analogue of this sample of \textit{JWST}'s LRDs - effectively a Big Red Dot at $z\sim2.5$.

Some SED models suggest that the photometry of these dusty AGN is best-fit by the combination of a dust-attenuated AGN and a massive star-forming galaxy \citep[e.g,][]{2024ApJ...968....4P}, whereas others prefer a scattered light scenario as in \ulas \citep[e.g.][]{2024ApJ...968...34W}. The inferred rest-frame UV luminosities of LRDs are significantly fainter than that of UV-selected AGN at similar epochs ($\rm -20 < M_{UV} < -16$), consistent with a less-than-one per cent scattering of the accretion disk emission, similar to \ulas \citep{2024ApJ...964...39G}. However, unlike the X-ray luminous \ulas, only 2/341 LRDs are detected in the X-ray. The two X-ray confirmed LRDs show evidence of relatively high column densities, $\rm log_{10}(N_H\, [cm^{-2}]) \sim 23$, consistent with \ulas and other HRQs. One interpretation for the absence of X-ray detections across the rest of the sample is that LRDs generally have denser gas columns with higher covering fractions than typically observed in HRQs, with the high column densities of neutral hydrogen extending to ISM scales \citep{2024arXiv240403576K}. Furthermore, \citet{2024ApJ...964...39G} propose that prolonged episodes of super-Eddington accretion could explain why LRDs are X-ray weak when compared to their optical luminosities. However, without large-scale spectroscopic confirmation it is still possible that some LRDs are starburst galaxies. Strikingly, the mid-infrared photometry of a sub-sample of LRDs show evidence of remarkably flat mid-infrared emission \citep{2024ApJ...968...34W}. One interpretation for the absence of hot (>1000K) dust emission is that these objects are not in-fact AGN, and instead the mid-infrared photometry is tracing the continuum from stellar populations whose spectrum peaks at $\sim 0.5-3 \mu m$ in the rest frame \citep{2002AJ....124.3050S}. However, since we also observe a similar flattening of the mid-infrared SED in \ulas and other HRQs, we propose an AGN obscured by material on ISM rather than nuclear scales, as an alternative explanation.

The similarities between HRQs and the LRDs are evident, however, the physical properties of these AGN differ from HRQs. LRDs are significantly less luminous than their cosmic-noon cousins, hosting black-hole masses $\sim10^{6-8}M_\odot yr^{-1}$ and $\lambda_{\rm Edd}\sim1$ \citep{2024arXiv240403576K}. Not only are they probing different regions of the quasar luminosity function, but also different regions of the M\textsc{bh}-$\rm \lambda_{Edd}$ plane. Despite this, it is clear that obscured accretion plays a major role in black-hole growth across the full dynamic range of black-hole mass and redshift.
 
\section{Conclusions} \label{SEC:Conc}

We have presented a high-resolution rest-frame UV to optical X-Shooter spectrum of a hyper-luminous heavily reddened quasar at $z=2.566$ - \ulas. We spectroscopically confirm the presence of excess UV emission relative to a dust-reddened quasar SED, which was initially noted in broadband photometric observations of this source. We fit the spectrum with several composite SED models to simultaneously explain the dust-attenuated rest-frame optical and excess rest-frame UV emission and also conduct a comprehensive analysis of the numerous emission and absorption line features detected in the spectrum. Our main findings are as follows: 

\begin{itemize}

\item {We confirm that the blue photometric colours of this source \citep{2018MNRAS.475.3682W} can not be explained by a simple dust-attenuated quasar model. If a secondary AGN component is responsible for the excess UV emission as seems likely based on the detection of broad emission lines in the UV, the luminosity of this blue AGN component is just 0.05 per cent of the total luminosity inferred from the dust-attenuated quasar component which has log$_{10}\{\lambda$L$_{\lambda} (3000\mathring{A}) [\rm erg\;s^{-1}]\}$ = 47.9 and a dust extinction of E(B-V)$^{\rm QSO}= 1.55$ mag. Based on analysis of the rest-frame UV imaging for this source, we rule out the presence of a dual AGN system. We conclude that we would require an extremely contrived dust geometry for just 0.05 per cent of the intrinsic AGN emission to "leak" through the obscuring medium and we therefore suggest that scattered AGN light scattering off dust in the interstellar medium of the host galaxy can explain the broad emission lines seen in the rest-frame UV and potentially the UV continuum emission.}

\item {We detect narrow [\ion{O}{ii}] emission without any significant emission from either [\ion{Ne}{iii}] or [\ion{Ne}{v}] and we therefore suggest that at least some of this [\ion{O}{ii}] emission is likely coming from the star-forming host galaxy of the dusty quasar. We also detect narrow Ly$\alpha$ emission at the systemic redshift of the quasar with an equivalent width and Ly$\alpha$ to NV ratio that is much higher than typically observed in blue quasars. Once again, some contribution from a star-forming host galaxy could be responsible for the narrow Ly$\alpha$ emission. Finally a composite dusty quasar + star-forming host galaxy fit to the X-Shooter spectrum provides a good match to the rest-frame UV and optical continuum shape though of course it cannot reproduce the broad emission lines in the UV. We conclude that the narrow low-ionisation nebular emission lines seen in the spectrum and potentially some of the UV continuum could also be coming from the quasar host galaxy. This is corroborated by analysis of the rest-frame UV image of this source from HyperSuprimeCam which shows evidence for spatially extended emission. }

\item {We analyse the mid infra-red SED of \ulas in an attempt to constrain the location of any dust that is responsible for attenuating the quasar emission and also potentially scattering some of the rest-frame UV emission to our line-of-sight. We find that \ulas has an extremely flat and atypical mid infrared SED with a ratio between the hot dust and accretion disk component at 2$\mu$m that is at least a factor of 5 lower than the average blue SDSS quasar at the same redshift. This suggests significant depletion of any hot ($>1000$K) dust component on toroidal scales. Hence, we propose that the bulk of the obscuring medium is on ISM, rather than nuclear, scales. An extended scattering medium is consistent with the blueshifts observed in the rest-frame UV high-ionisation lines -e.g. \ion{C}{iv} and \ion{N}{v}, since a dense circumnuclear cocoon would likely result in the symmetrical line profiles such as those observed in ERQs. }

\item We detect numerous associated absorption line (AAL) systems in the rest-frame UV across a range of species and velocity components including blueshifted absorption features in \ion{C}{iv}, \ion{N}{v}, \ion{Si}{iv} and \ion{Si}{iii} as well as potentially redshifted features in \ion{Si}{iii} and \ion{C}{iv}.  

\item {Using the Balmer lines and the optical continuum luminosity ($\rm L_\lambda(5100\,$\AA)), we estimate the black-hole mass $\rm log_{10}(H\beta, M\textsc{bh} [M_{\odot}]) = 10.26 \pm 0.05$, bolometric luminosity L$_{\rm Bol}$ = $ \rm 10^{48.16}\, erg\;s^{-1}$ and Eddington-scaled accretion rate log$_{10}$($\rm \lambda_{Edd}) = -0.19$. Our optical-derived bolometric luminosity implies an X-ray bolometric correction $\rm log_{10}(k_{Bol},L_X) = 2.56$. The bolometric correction implied from the mid-infrared luminosity $\rm log_{10}(k_{Bol},L_{6\mu m}) = -0.66$.}

\item {We find evidence for significant outflows affecting both the BLR and NLR gas. The \ion{C}{iv} blueshift is 1080$\pm$110 $\rm km\,s^{-1}$ and the [\ion{O}{iii}] W80 is 1830$\pm$350 $\rm km\,s^{-1}$. However, in the context of outflows seen in other luminous quasar populations, these values are not extreme and could be explained by the significant black hole mass of \ulas. Using the [\ion{O}{iii}] velocity we estimate the mass outflow rate for the ionised gas $\dot M_{ion}^{out} = 199 \,M_\odot\, \rm yr^{-1}$, with a corresponding kinetic power $\dot \epsilon_{k}^{ion} = 10^{44.61} \rm erg\,s^{-1} \sim 0.001\,L_{Bol}$. This is consistent with SDSS blue quasars at comparable luminosity. }

\item {Despite probing a different region of the quasar luminosity function and M\textsc{bh}-$\rm \lambda_{Edd}$ plane, we find that \ulas represents a cosmic-noon analogue of the \emph{JWST} LRDs. Both LRDs and HRQs exhibit blue rest-frame UV photometric colours, which are interpreted as some combination of scattered AGN and star-forming host galaxy emission. As both populations show signs of hot dust depletion in their mid-infrared SEDs, we propose that the obscuring medium in both populations is likely on galaxy-wide rather than torus scales.}

Over the coming years, large spectroscopic surveys in the near infra-red e.g. with VLT-MOONS \citep{Maiolino:20} and the Prime Focus Spectrograph (PFS; \citealt{Greene:22}) will offer the opportunity to characterise the properties of many more red AGN effectively bridging the gap in both luminosity and redshift between HRQs such as \ulas and the \textit{JWST} LRD population.  

\end{itemize}

\section*{Acknowledgements}

We thank Jenny Greene for a constructive referee's report and Christian Knigge and James Matthews for their invaluable insight, which helped shape the discussion. We also thank J. Xavier Prochaska and the PypeIT user community for advice on using PypeIT during the data reduction stages of this work. 

MS acknowledges funding from the University of Southampton via the Mayflower studentship. MB, ST and SM acknowledge funding from the Royal Society via a University Research Fellowship Renewal Grant and Research Fellows Enhancement Award (URF/R/221103; RF/ERE/221053). MT acknowledges funding from the Science and Technology Funding Council (STFC) grant, ST/X001075/1.

\section*{Data Availability} 

The X-Shooter data presented in this paper can be downloaded from the European Southern Observatory's (ESO) online  archive\footnote{\url{http://archive.eso.org/eso/eso_archive_main.html}}, with ESO Program ID: 099.A-0755(A).

\bibliographystyle{mnras}
\bibliography{Citations} 

\appendix 

\section{PypeIt reduction parameters} \label{APP:PypeIt_User_Params}

For reproduciblity, we detail the PypeIt user-level parameters that were changed from their default values to reduce the raw spectrum of \ulas in Table \ref{tab:PypeIt_User_Params}. 

\begin{table}
    \centering
    \caption{We present the various PypeIt user-level parameters used to optimise the reduction and extraction of the \ulas X-Shooter spectrum. The numerous additional parameters, which do not feature in this table, were left at their default settings.}
    
    \begin{tabular}{|l|l|l|}
         \hline
         \hline
          UVB/VIS & &\\
         \hline
         \hline
         Primary block & Secondary Block & Parameter \\
         \hline
          reduce & findobj & snr$\_$thresh = 3.0 \\
          reduce & findobj & ech$\_$find$\_$max$\_$snr = 1.0 \\
          reduce & findobj & ech$\_$find$\_$min$\_$snr = 0.1 \\   
          reduce & findobj & ech$\_$find$\_$nabove$\_$min$\_$snr = 2.0 \\ 
          calibrations & wavelengths & rms$\_$threshold = 2.0 \\
          calibrations & tilts & sig$\_$neigh = 5.0 \\
          calibrations & tilts & tracethresh = 5.0 \\
          calibrations & slitedges & tracethresh = 7.0 \\
         \hline
         \hline
          NIR & & \\
         \hline
         \hline
         Primary block & Secondary Block & Parameter \\
         \hline         
          reduce & N/A & trim$\_$edge = 1,1 \\         
          reduce & findobj & find$\_$trim$\_$edge = 1,1 \\
          reduce & findobj & snr$\_$thresh = 5.0 \\  
          reduce & findobj & ech$\_$find$\_$max$\_$snr = 1.0 \\
          reduce & findobj & ech$\_$find$\_$min$\_$snr = 0.1 \\   
          reduce & findobj & ech$\_$find$\_$nabove$\_$min$\_$snr = 2.0 \\ 
          reduce & skysub & no$\_$local$\_$sky = True \\  
          reduce & skysub & global$\_$sky$\_$std = False \\  
          reduce & extraction & use$\_$2dmodel$\_$mask = False \\  
          calibrations & wavelengths & rms$\_$threshold = 1.1 \\
          calibrations & tilts & sig$\_$neigh = 5.0 \\
          calibrations & tilts & tracethresh = 5.0 \\
          calibrations & slitedges & tracethresh = 7.0 \\
         \hline
         \hline    
    \end{tabular}
    \label{tab:PypeIt_User_Params}
\end{table}

\section{Degeneracies in the reddened quasar + blue quasar light SED model} \label{APP:SED_Corner}

\begin{figure}
 \includegraphics[width=\linewidth]{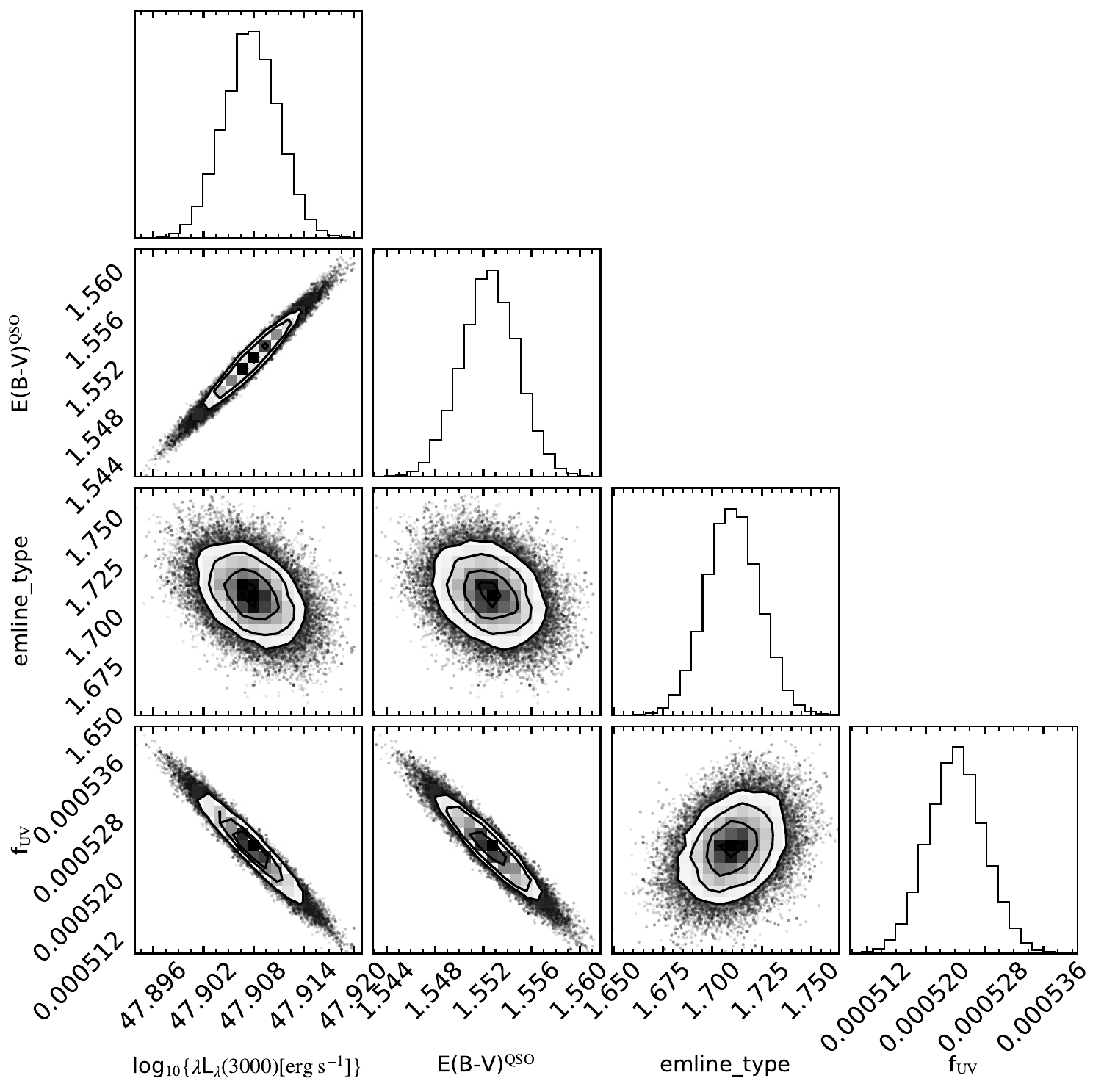} 
  \caption{The reddened quasar + blue quasar light SED model (see Section \ref{SEC:SED_Fit}) corner plots. In order left-to-right/top-to-bottom the marginalised distributions represent log$_{10}\{\lambda$L$_{\lambda} (3000\mathring{A})\}$, E(B-V)$^{\rm QSO}$, $\rm emline\_type$ and f$_{\rm UV}$. The corner plot illustrates that the model is well-converged and highlights the degeneracies between several model parameters.}
 \label{fig:SED_Corner}
\end{figure}

Here we discuss the degeneracies in the model parameters associated with the first SED fit detailed in Section \ref{SEC:SED_Fit}. We present the corner plot for this SED model in Fig. \ref{fig:SED_Corner}. The marginalised distributions show that each parameter is well converged, however, Fig. \ref{fig:SED_Corner} reveals several degeneracies. We observe a strong anti-correlation between f$_{\rm UV}$ and both log$_{10}\{\lambda$L$_{\lambda} (3000\mathring{A})\}$ and E(B-V)$^{\rm QSO}$ which in turn are positively correlated with each other. This behaviour is expected as should the quasar be more luminous, a lower scattering fraction is required to reproduce the rest-frame UV emission. Equally, should the quasar suffer heavier dust attenuation, we require a more luminous quasar to reproduce the rest-frame optical and hence the scattering fraction must again be lower to compensate for the increase in luminosity. The $\rm emline\_type$ parameter is independent of the other three model parameters since the \citet{2021MNRAS.508..737T} SED model allows for line strength variations independent of the continuum parameters.

\bsp	% typesetting comment
\label{lastpage}
\end{document}